\documentclass[8.5pt]{article}
\usepackage[super,sort&compress,comma]{natbib} 
\usepackage{mhchem}
\usepackage{times,mathptmx}
\usepackage{sectsty}
\usepackage{balance}

\usepackage{graphicx} 
\usepackage{epstopdf}
\usepackage{lastpage}
\usepackage[format=plain,justification=raggedright,singlelinecheck=false,font=small,labelfont=bf,labelsep=space]{caption} 
\usepackage{fancyhdr}
\begin{document}

\thispagestyle{plain}
\fancypagestyle{plain}{
\renewcommand{\headrulewidth}{1pt}}
\renewcommand{\thefootnote}{\fnsymbol{footnote}}
\renewcommand\footnoterule{\vspace*{1pt}%
\hrule width 3.4in height 0.4pt \vspace*{5pt}} 
\setcounter{secnumdepth}{5}

\makeatletter 
\def\subsubsection{\@startsection{subsubsection}{3}{10pt}{-1.25ex plus -1ex minus -.1ex}{0ex plus 0ex}{\normalsize\bf}} 
\def\paragraph{\@startsection{paragraph}{4}{10pt}{-1.25ex plus -1ex minus -.1ex}{0ex plus 0ex}{\normalsize\textit}} 
\renewcommand\@biblabel[1]{#1} 
\renewcommand\@makefntext[1]%
{\noindent\makebox[0pt][r]{\@thefnmark\,}#1}
\makeatother 
\renewcommand{\figurename}{\small{Fig.}~}
\sectionfont{\large}
\subsectionfont{\normalsize} 

\fancyfoot{}
\fancyfoot[RO]{\footnotesize{\sffamily{1--\pageref{LastPage} ~\textbar \hspace{2pt}\thepage}}}
\fancyfoot[LE]{\footnotesize{\sffamily{\thepage~\textbar\hspace{3.45cm} 1--\pageref{LastPage}}}}
\fancyhead{}
\renewcommand{\headrulewidth}{1pt} 
\renewcommand{\footrulewidth}{1pt}
\setlength{\arrayrulewidth}{1pt}
\setlength{\columnsep}{6.5mm}
\setlength\bibsep{1pt}

\twocolumn[
 \begin{@twocolumnfalse}
\noindent\LARGE{\textbf{Density Functional Theory Evaluation of Cation-doped Bismuth Molybdenum Oxide Photocatalysts for Nitrogen Fixation}}
\vspace{0.6cm}

\noindent\large{\textbf{Alhassan S. Yasin\textit{$^{a}$} and Botong Liu\textit{$^{a}$} and Nianqiang Wu\textit{$^{a}$} and Terence Musho,$^{\ast}$\textit{$^{a}$}}}\vspace{0.5cm}

\noindent\textit{\small{\textbf{Received Xth XXXXXXXXXX 20XX, Accepted Xth XXXXXXXXX 20XX\newline
First published on the web Xth XXXXXXXXXX 200X}}}

\vspace{0.6cm}

\textbf{Abstract}\\
\noindent \normalsize{This study investigates the photocatalytic nitrogen fixation on a cation-doped surface (Bi$_{x}$M$_{y}$)$_2$MoO$_6$ where (M = Fe, La, Yb) in both the orthorhombic and monoclinic configurations using a density functional theory (DFT) approach with experimentally validated model inputs. The proceeding discussion focuses on the Heyrovsky-type reactions for both the associative and dissociative reaction pathway related to nitrogen reduction. Key fundamental insight in the reduction mechanism is discussed that relates the material properties of the substitutional ions to the nitrogen and hydrogen affinities. Physical insight is gathered through interpretation of bound electronic states at the surface. Compositional phases of higher Fe and Yb concentrations resulted in decreased Mo-O binding and increased affinity between Mo and the N and H species on the surface. The modulation of the Mo-O binding is induced by strain as Yb and Fe are implemented, this, in turn, shifts energy levels and modulates the band gap energy by approximately 0.2 eV.  This modification of Mo-O bond as substitution occurs is a result of the orbital hybridization of M-O  (M = Fe, Yb) that causes a strong orbital interaction that shifts states up toward the Fermi. The optimal composition was predicted to be an orthorhombic configuration of (Bi$_{0.75}$Fe$_{0.25}$)$_2$MoO$_6$ with a predicted maximum thermodynamic energy barrier of 1.4 eV. This composition demonstrates effective nitrogen and hydrogen affinity that follows the associative or biological nitrogen fixation pathway.}
\vspace{0.5cm}
 \end{@twocolumnfalse}
 ]

\footnotetext{\textit{$^{a}$~Department of Mechanical and Aerospace Engineering, West Virginia University, Morgantown, WV 26506-6106, USA. Fax: 304-293-6689; Tel: 304-293-3256; E-mail: tdmusho@mail.wvu.edu}}

\section{Introduction}
Biological development and adaptation on this planet is possible due to essential processes of converting nitrogen into ammonia, referred to as nitrogen fixation.~\cite{hoffman14,ferguson98,burris01,canfield10} This essential process allows the production of reduced forms of nitrogen that living organisms require for sustainability. These organisms utilize the reduced forms of nitrogen to activate an enzyme-catalyzed process where substances are converted into more complex products for basic building blocks~\cite{hoffman14}, referred to as biogenesis~\cite{burris01}. Nitrogen is one of the most abundant elements on Earth, dominantly as nitrogen gas (N$_{2}$) in the atmosphere~\cite{jia14,mackay04}. However, this form of nitrogen needs to be reduced in order for meaningful utilization by living organisms, thus taking forms through: (i) ammonia (NH$_{3}$) and/or nitrate (NO$_{3}^{-}$) fertilizer~\cite{canfield10,thamdrup12}, (ii) utilization of released compounds during organic matter decomposition~\cite{hoffman14,burris01}, (iii) the conversion of atmospheric nitrogen by natural processes, such as lightening~\cite{gruber08}, and (iv) biological nitrogen fixation (BNF)~\cite{hoffman14}. 

Semiconductor photocatalysts have been gaining greater attention for a broad range of processes due to the growing concerns about the environmental pollution and future demands. Making the design process of a new semiconducting photocatalysts is that they multifaceted requirement that these new materials must optimize several attributes simultaneously, such as high selectivity, low cost, electrochemical stability, and be environmentally friendly~\cite{lai12}. More recently researchers have been exploring active oxide semiconducting photocatalysts due to their large design space and sensitivity to the visible light spectrum. An example of these oxide types photocatalyst include SrTiO$_3$, ZrO$_2$, BaTi$_4$O$_9$, bismuth oxides, etc.~\cite{lai12}. Compound this with the fact that oxides typically exhibit good thermal stability and inherently low reactivity, makes them good material candidates where the cation concentration can be controlled to modulate the reactivity and selectivity. The aim of this study is to demonstrate key properties that characterize BNF processes for a Bismuth-based photocatalysts (Bi$_2$MoO$_6$), which have attained much attention due to the up-shifted valence band and a narrow band gap of 2.63 ev~\cite{kumari15} that has the ability to utilize more sunlight. This small band gap is a  result of hybridization between Bi 6s and O 2p states~\cite{lai12}, and this hybridization makes the valence band more disperse which allows more easily transited photogenerated holes in the valence band. The system of interest for this study is based on Bi$_2$MoO$_6$ of an orthorhombic crystal structure and its counterpart Bi$_3$FeMo$_2$O$_{12}$ of a monoclinic crystal structure with the substitution of Bi and Fe sites with combinations of Fe, La, and Yb.

\subsection{Biological Nitrogen Fixation (BNF)}
Biological nitrogen fixation (BNF) process occurs naturally due to nitrogen-fixing bacteria in the soil, these bacteria are associated with a specific groups of plants that utilize living organisms called prokaryotes to fix the nitrogen in the soil. These organisms are very reliant on the nitrogenases~\cite{raymond04,hoffman14}, which are enzymes that can be produced by certain bacteria, and are responsible for the reduction of nitrogen to ammonia. Atmospheric nitrogen, which takes the form of N$_{2}$ at atmospheric temperature and pressure is comprised of two nitrogen atoms joined by a triple covalent bond. The stability of the dinitrogen molecule is correlated with the triple covalent bond and manifest inself in a fairly inert~\cite{jia14,mackay04} and nonreactive molecular state. Nature has found an efficient method to dissociate this bond using a efficient reaction pathway that is targeted in the proposed research. More specifically, microorganisms that fix nitrogen (nitrogenases) require 16 moles of adenosine triphosphate (ATP) to reduce each mole of nitrogen (approximately 5 eV)~\cite{hubbell98}. Typically the reduction of atmospheric nitrogen by nitrogenases occurs by initially weakening the N-N bond by successive protonation until the dissociation barrier is low enough that the N-N dissociates (later in the reaction sequence)~\cite{howalt13}. This barrier to overcome is referred to as the positive determinant step which requires the most energy to evolve the reaction. This particular reaction sequence is referred to as the associative mechanism, and the energy required by the microorganisms to reduce nitrogen is obtained by oxidizing surrounding organic molecules. In other words, microorganisms that are non-photosynthetic must obtain this energy from other organisms, while photosynthetic capable microorganisms use sugars produced by photosynthesis to obtain the essential energy to oxidize other organic molecules~\cite{hubbell98,buttel98}.

In taking a cue from nature and the evolution of the nitrogenase enzyme, it is found that the active site in the nitrogenase enzyme is a cluster of FeMo$_{7}$S$_{9}$N, the FeMo-cofactor, with an electrochemical reaction,
$N_{2}+8(H^{+}+e^{-}) \rightarrow 2NH_{3}+H_{2}$.
These clusters have shown great stability in various configurations and sustainability in natural environments and is often found as a trace element in soils. Thus, most synthetically researched nitrogen fixation processes of transition metal configurations are based on some variation to molybdenum (Mo) structures.

\subsection{Haber-Bosch Process}
The most known method of synthetically producing ammonia in large quantities for the modern world is through the utilization and production of the Haber-Bosch process~\cite{gruber08,hoffman14}. Fundamentally this process reduces nitrogen the same way as natural biological systems~\cite{vitousek97}, but with a required energy substantially greater than that of biological processes. The way this process converts atmospheric nitrogen (N$_{2}$) to ammonia (NH$_{3}$) is by successive reactions of nitrogen gas with hydrogen gas (H$_{2}$), coupled with metal catalysts (usually an iron-based catalyst). However, this synthetic process requires large amounts of energy and utilization of hydrogen, which is typically obtained by processing natural gas in large quantities that cause carbon dioxide emission (CO$_{2}$)~\cite{vitousek97,hoffman14}.

This reaction of nitrogen and hydrogen gas takes place under high temperature and pressure~\cite{burgess96,eady96}, where some excess energy can be recovered and used to continue the manufacturing process, this is referred to as recuperation. During the process N and H, gas molecules are heated to approximately 400 to 450 $^{\circ}$C, while continuously being kept at a pressure of 150 to 200 atm~\cite{burgess96}. At this point of the manufacturing process, it is necessary to remove as much of the surrounding oxygen as possible prior to the exposure of gases to the catalyst to avoid catalytic oxidation. After the oxygen removing step is done, the dissociated nitrogen and hydrogen mixture is passed over a Fe-based catalyst to form ammonia.
\begin{equation}
N_{2}(g)+ 3H_{2}(g) \rightleftharpoons 2NH_{3}(g)
\label{equ:1}
\end{equation}

Contrary to biological nitrogen fixation where the N$_{2}$ bond is broken late in the reaction sequence after initially weakening the bond by successive protonation, the Haber-Bosch process initially dissociates the N$_{2}$ bond on the first step and then protonates each nitrogen atom, referred to as the dissociative mechanism~\cite{howalt13}. Nitrogen and hydrogen atoms do not react until the strong N$_{2}$ triple bond and H$_{2}$ bond have been broken~\cite{skulason12} for the dissociative mechanism. Even though this reaction is reversible and an exothermic process, relatively high temperature and pressure are needed still to allow the reaction evolve quickly~\cite{kozuch08}. However, this high temperature shifts the equilibrium point towards the reactants side thus resulting in the lower conversion of ammonia~\cite{skulason12,howalt13}. To correct for reactants equilibrium shift, external energy is thus needed to shift the equilibrium in favor of the reactions products~\cite{skulason12}.

\begin{figure}[!h]
\begin{center}
\textbf{\begin{Large}A\end{Large}}\includegraphics[width=.7\columnwidth]{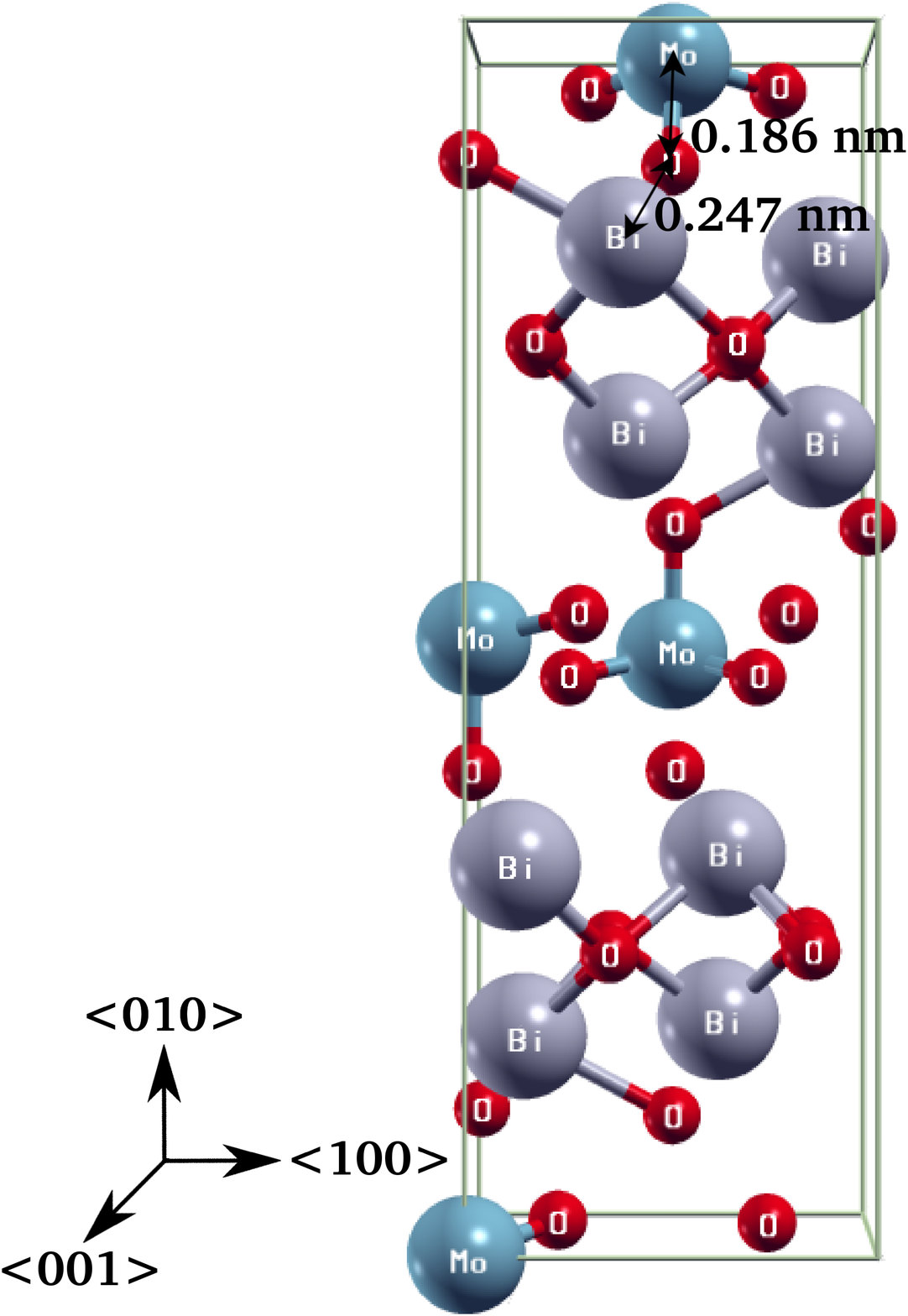} \\
\textbf{\begin{Large}B\end{Large}}\includegraphics[width=.7\columnwidth]{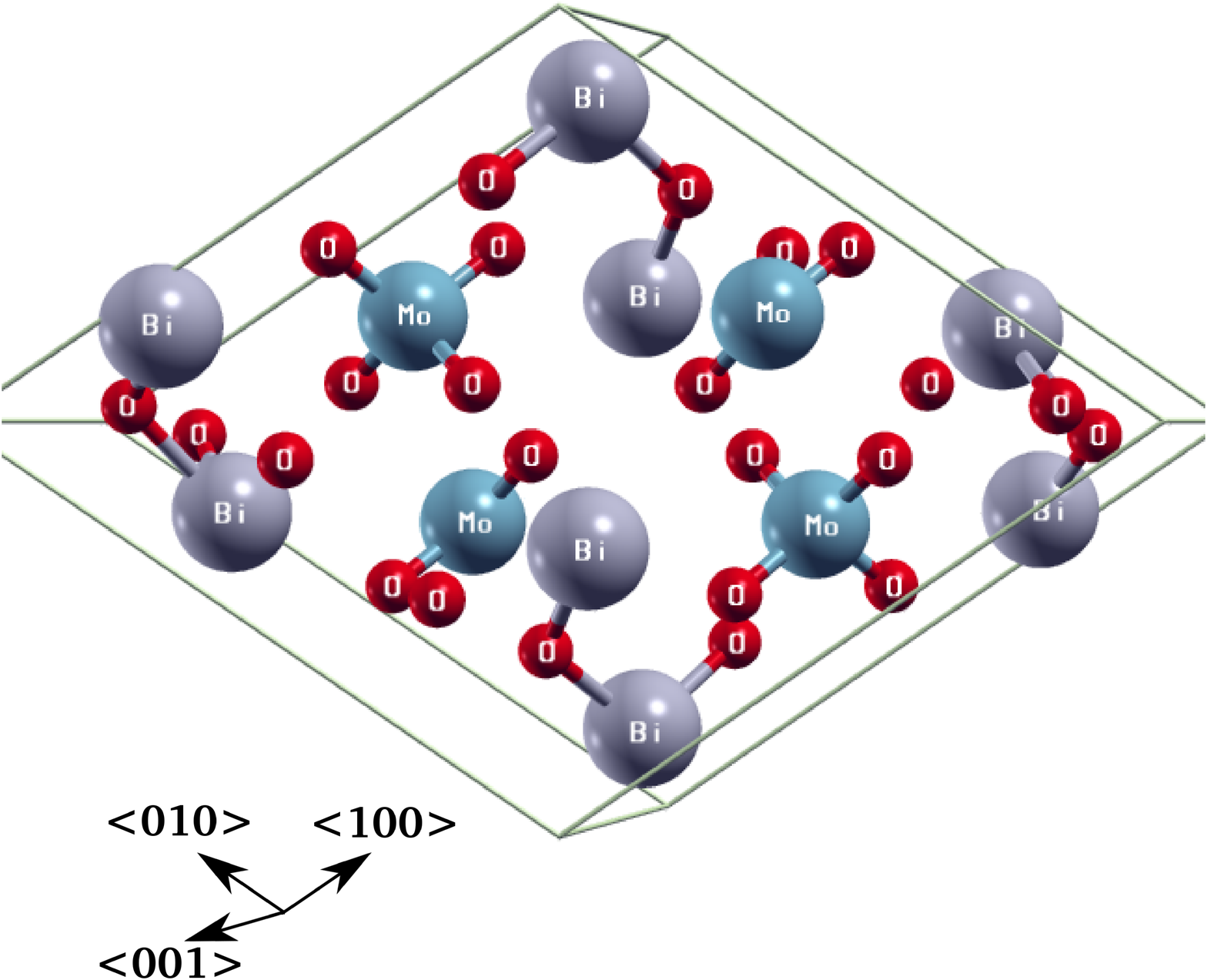}
\end{center}
\caption{Illustration of the Bi$_2$MoO$_6$ in orthorhombic and monoclinic configurations. Sub-figure A is orthorhombic unit cell made up of 24 O atoms 8 transition metal atoms (Bi) and 4 Mo atoms, sub-figure B is monoclinic unit cell of the same ratio number of atoms making a total of 36 atoms in each configuration. Fe, La, and Yb are subed into the 8 metal atom positions (Bi) for both structures. The 10 {\r{A}} vacuum in A is placed along the [010] plane and in B is placed along the [100] plane. Indicated distances for sub-figure A shows the internal bond length between M-O-Bi.}\hrule
\label{fig:UC}
\end{figure}

\section{Methodology}
This study aims to understand is the potential utilizing concepts of electrochemistry that can aide to further investigate synthetically production ammonia. This is enhanced by the demand for ammonia to sustain the very apparent population influx. Great strides have been made to address the obstacles faced while developing catalytic materials for nitrogen reduction in the past several years from a theoretical~\cite{howalt13,raymond04,hoffman14} and experimental~\cite{howalt13,heyrovsky27,tafel05} perspective, which gives a starting point for further investigation on these catalytic materials. In fact, some of these studies showed that ammonia synthesis to be very structure sensitive on metal surfaces that can occur on surface steps of Fe and Ru~\cite{howalt13}, and potently could expect the associative mechanism to be even more so. Thus, bringing about our previous study to investigate the undercoordinated MoS$_{2}$ structure and current study of Bismuth containing structure, which provides a means to investigate this structure sensitive associative mechanism. 

The utility of density functional theory calculations allows for a first principle prediction of the chemical reactions of the Bismuth containing structure for photocatalytic ammonia production. This study focuses on the Bismuth containing structures of Bi$_2$ MoO$_6$ which has reported to possesses intrinsic activation ability for nitrogen-splitting~\citep{hao16}. The edge exposed coordinatively unsaturated Mo atoms can effectively achieve activation, chemisorption, and photo-reduction of dinitrogen, because Mo atoms act as a nitrogen activation center in the Mo-O coordination polyhedron~\citep{hao16,lai12}. On the same basis, Bi$_2$MoO$_6$ based artificial photosynthesis system was successfully constructed by controllable crystal structure and defect engineering~\citep{hao16,kumari15}. The result was effective energy coupling with photons, excitons, and di-nitrogen in this system which demonstrates impeccable sunlight-driven nitrogen fixation performance~\citep{hao16}.

\subsection{Photocatalytic Reduction of Nitrogen}
The main concern of successfully reducing nitrogen by synthetic means is the energy required to implement these processes. Usually, the main source of energy utilized by these synthetic processes is attained by burning fossil fuels that yield environmental pollution as discussed in previous sections. This prompted studies to investigate and advance better reaction pathways for ammonia production through the mimicry of biological organisms, thus allowing various researchers to develop catalysts for this process. Thus allowing many researchers to investigate the mechanism of nitrogen fixation and produce photocatalytic and photoelectrochemical processes to utilize solar energy conversion. 

The idea is to allow the presence of a catalyst to accelerate the process of photo-reactions, this is because the photocatalytic activity is dependent on the catalyst to create hole pairs (electron charge carriers), which generate free unpaired valence electrons that can contribute to secondary reactions. Typically, these photocatalysts are comprised of transition metal oxides and or semiconductors that have void energy region (band gap) where no energy levels are present to undergo an electron and hole recombination produced by photoactivation (light/photon absorption). When the energy of an absorbed photon is equal to or greater than that of the materials band gap, an electron becomes excited from the valence band (HOMO) to the conduction band (LUMO), thus producing a positive hole in the valence band. This excited electron and hole can then become recombined, and release heat from the energy gained by the photon exciting the surface of the material. The idea is to have a reaction that produces an oxidized product by reaction of generated holes with a reducing agent and to have a reduced product by the interaction of the excited electron and an oxidant. Simply-put, photoelectrochemical and photocatalytic processes rely on semiconductor materials to absorb sunlight and generate excited charge carriers that allow for reactions to take effect without any external input.

Thus the presented work aims to yield a photocatalytic and electrochemical investigation for sunlight-driven nitrogen fixation to reduce energy consumption, that will also provide flexibility in designing materials for nitrogen conversion. This work prompts to gain a crucial understanding of nitrogen to ammonia reaction process on the photocatalytic and electrochemical level and to develop means of approaching a sustainable solar driven conversion process of nitrogen to ammonia.

\subsection{Material Design Space of Interest}
Researchers have previously studied faceted metal surfaces of the Bi$_2$ MoO$_6$ orthorhombic structure, and limited research has investigated the Bi$_3$FeMo$_2$O$_{12}$ monoclinic structure. The following study has performed calculations on a basis for configuration of (Bi$_{x}$M$_{y}$)$_2$MoO$_6$ where (M = Fe, La, Yb) sub-configurations of both the orthorhombic and monoclinic structures. It is noted that ratios of the expressed systems are the same; however, the Bi$_2$MoO$_6$ based structures are of orthorhombic crystal structure and Bi$_3$FeMo$_2$O$_{12}$ based structures are of the monoclinic crystal structure. This permits investigations of reaction intermediates for both the dissociative and associative mechanisms on Bismuth containing structures for Fe, La, and Yb transition metals, and also allows the study of stepped and closed packed metal surface relations. The expressed analyses of this paper utilize fundamental calculations to construct Gibbs free energy diagrams for both the dissociative and associative reaction pathways. These diagrams will be used to determine the lowest potential barrier across the investigated metal configurations, this will be the barrier required to overcome in each configuration to evolve the reaction.

The material of interest is Bi$_2$MoO$_6$ shown in Figure~\ref{fig:UC} for the orthorhombic and monoclinic configuration has received tremendous attention due to the attractive optical properties, catalytic properties, electronic properties, and its high chemical stability. Bi$_2$MoO$_6$ has been identified as a promising photocatalytic material due to Aurivillius layered structure comprised of MoO$_6$ octahedron and Bi-O-Bi layers as shown in Figure~\ref{fig:UC}A and B.\citep{lai12,kumari15,nie16} Aurivillius is a form of perovskite structure that is built by alternating layers of (Bi$_2$O$_2$)$^{2+}$ and pseudo perovskite blocks and these perovskite layers are 'n' octahedral layers in thickness, thus making (Bi$_2$O$_2$)$^{2+}$ layers in-between MoO$_6$ octahedral layers that are connected to each other at corners. Electron-hole pairs will migrate easily and transfer effectively to the surface of a corner-sharing structure, due to the contribution of photocatalytic performance under visible light.\citep{hao16,nie16} Experimental observations have approximated the band gap of Bi$_2$MoO$_6$ to 2.56-2.63 eV,~\citep{lai12,weng16,hao16,kumari15,nie16} with unique physical and chemical properties, such as environmentally friendly, chemical inertness and photo-stability features, making them ideal photocatalytic materials.

Bi$_2$MoO$_6$ structure is unique in it forms a layered structure where individual layers are primarily held together with electrostatic forces. The main structural modification that was done is the substitution of the non-organic ion Bi with various combinations in all possible positions in the unit cell of (BiM)$_2$MoO$_6$ (M = Yb, Fe, La). It is noted that calculations are done for Bi$_2$MoO$_6$ based structures in both the orthorhombic Figure~\ref{fig:UC}A and monoclinic Figure~\ref{fig:UC}B configuration. It is further noted that for BNF Mo-O plays an important role in attracting nitrogen to the surface which proves to be ideal for nitrogen fixation (Mo acts as an N activation center in the Mo-O coordination~\citep{hao16,lai12}), thus even though the monoclinic configurations of Bi$_2$MoO$_6$ (Figure~\ref{fig:UC}B) demonstrate similar properties as the orthorhombic configuration (Figure~\ref{fig:UC}A), its still theorized that monoclinic structure will not yield dramatically better results~\cite{nie16}. This is because the Mo-containing complexes are located in the center of the monoclinic structure Figure~\ref{fig:UC}B, thus making the recombination between the holes and electrons to be weak~\cite{nie16}, and ideally for nitrogen fixation, the interaction of Mo and N species are of most importance. Looking at orthorhombic configuration, Figure~\ref{fig:UC}A, the first layer N species will see in this configuration is the aliened Mo atoms, reference Figure~\ref{fig:ads}, that prove to better stabilize N species than that of monoclinic configuration. Never the less the monoclinic configuration is still looked at for comparability and data collection.

\begin{figure}[!ht]
\includegraphics[width=1.0\columnwidth,trim=3.5cm 0cm 3.0cm 1cm, clip=true]{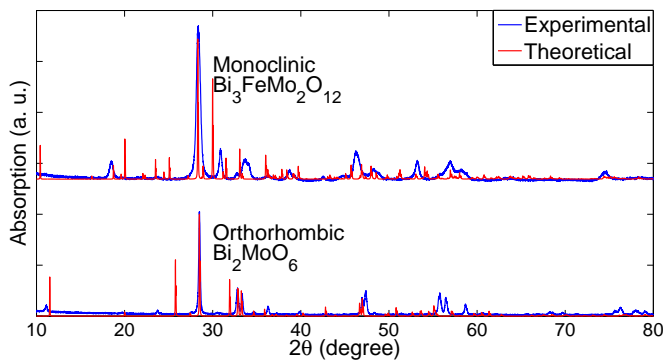}
\caption{Illustration of the diffraction patterns for Bi$_2$MoO$_6$ in orthorhombic and Bi$_3$FeMo$_2$O$_{12}$ in monoclinic configurations of the experimental and theoretical calculations. The theoretical calculations are of T = 0 K for ideal crystalline unit cells, were as experimental is done at room temperature for polycrystalline unit cells which tend to have some impurities, thus resulting in some variations in diffraction pattern.}\hrule
\label{fig:xrd}
\end{figure}

To further validate calculations on the two phase of orthorhombic structure (that is based on Bi$_2$MoO$_6$) and monoclinic structure (that is based on Bi$_3$FeMo$_2$O$_{12}$), analyzes of powder diffraction patterns are analyzed to ascertain the calculations done for the reaction pathways for the correct configurations. Figure~\ref{fig:xrd} expresses the XRD pattern of the two configurations calculated for the theoretical and experimental results. The theoretical calculations are for unit-cells of relaxed structure where as the experimental is for polycrystalline unit-cells that may contain impurity states. This means that one would expect the experimental to produce broader peaks that may result from the resolution and combination of various peaks detected by the equipment and that the theoretical calculations will result in sharper peaks, which is true of Figure~\ref{fig:xrd}. It is noted that peak positions are aliened in Figure~\ref{fig:xrd}, which suggests that calculations done on the expressed configurations in Figure~\ref{fig:UC} are consistent with the experimental configurations analyzed.

\subsection{Material Design Strategies} 
A good strategy to design new catalyst for nitrogen reduction is by utilizing components that lie from both the dissociative and associative side. This will allow structures to be designed for more effective breaking of the N-N bond by allowing good ratios of N and H species affinity, where associative protonation requires better H affinity where as the initial braking of N-N for the dissociative will require more N affinity. Initial DFT calculations express Bi$_2$MoO$_6$ orthorhombic and Bi$_3$FeMo$_2$O$_{12}$ monoclinic structures to favor the dissociative side and with the addition of Fe, La, and Yb, which can favor the associative mechanism, the material of interest takes the form of (Bi$_x$M$_y$)$_2$MoO$_6$ (M = Yb, Fe, La) for orthorhombic and monoclinic configurations. Main structural modification that was conducted is the substitution of the Bi atoms with various combinations of the 8 possible positions in the unit cell Figure~\ref{fig:UC}A and B. The 8 atomic positions of both orthorhombic and monoclinic structures that were modified are clearly shown in Figure~\ref{fig:UC} of the unit cell. All element configuration combination of the three elements (Yb, Fe, La) is implemented and tested in the metal positions expressed in Figure~\ref{fig:UC} of the unit cell. These configurations are classified under one element case (pure Bi, Yb, Fe, La structure) and two element cases (combination of Bi and Yb, Bi and Fe, Bi and La). 

It is noted that while the Bi$_2$MoO$_6$ material is subbed with Fe, La, and Yb, combination of the two elements are evaluated in this study to better vary the mechanism of interest for each distinct adsorption species of N$_{2}$H$_{x}$ and NH$_{x}$ respectably. Of the two element case combinations, there are different configurations possible for same elemental ratios. Meaning the minority ion can be placed in several substitutional locations. This makes a total of 768 (2 elements for 8 positions for 3 structures 2$^{8}$x3 of Bi and La, Bi and Yb, Bi and Fe) unique configurations for orthorhombic and 768 configurations for monoclinic structures. Only the configurations that yielded the lowest energy of the relaxed case for each distinct structure will be evaluated for the electrochemical reactions. For example, configuration of 1 atom Bi and 7 atom Fe (Bi1Fe7) can have 8 distinct configurations of Bi being in any of the 8 metal positions, so only the lowest energy structure after the relaxation will be implemented in the electrochemical reactions, and this is the same for every other configuration for Bi$_2$MoO$_6$ based orthorhombic and monoclinic structures (Bi2Fe6, Bi3Yb5, Bi4La4, ...). Each configuration also had various operations that must be done in sequence to obtain proper data, like initial relaxation, screening of best configurations, then vacuum implementation and relaxation. 

This resulted in a total of 27 configurations for orthorhombic and 27 configurations for monoclinic structures that had the lowest energy structure for a specific configuration. The reader should note that this is a tremendous amount of computational and analytically challenging process. The brute force approach allows every configuration of the orthorhombic and monoclinic structures to be explored and give the best configuration of (Bi$_x$M$_y$)$_2$MoO$_6$ (M = Fe, La, Yb) structure as a means of nitrogen reduction catalyst.

\subsection{Associative and Dissociative Reaction Pathways}
In the electrochemical process of forming ammonia, it is ideal to reference the source of protons and electrons to model the anode reaction,
\begin{equation}
H_{2} \rightleftharpoons 2(H^{+}+e^{-}).
\label{equ:2}
\end{equation}
Initially, protons are exposed into the proton conducting electrolyte to maintain the equilibrium and diffuse into the cathode while an external circuit is used to transport electrons to the cathode side through the substrate. At this instant, a nitrogen molecule will react with surrounding protons and electrons at the cathode as described by the following reaction,
\begin{equation}
N_{2}+6(H^{+}+e^{-}) \rightarrow 2NH_{3},
\label{equ:3}
\end{equation}
to form ammonia at the catalytically active site. This is known as the overall electrochemical reaction.

In theory, the reaction can take two different possible pathway forms in which there are two different possible types of mechanism for each pathway in order to synthesize ammonia electrochemically. Pathways correspond to either the Tafel-type mechanism or the Heyrovsky-type mechanism.~\cite{heyrovsky27} In the Tafel-type mechanism, solvated protons from the solution are the first adsorb~\cite{tafel05} on the surface and combine with electrons, then the hydrogen adatoms react with the adsorbed species(N$_{2}$H$_{x}$ or NH$_{x}$). Indirect effect through interchangeable concentrations of the reactants can only be done by this type of mechanism.~\cite{heyrovsky27} It is noted that this study focuses on room temperature processes, and that activation barriers for Tafel-type reactions are about 1 eV or higher for most transition metal surfaces~\cite{honkala05,wang11}, thus most likely yielding very slow reactions for this type of mechanism. Also, it is further noted that this type of mechanism is known to require hydrogenation steps~\cite{wang11} of the reaction barriers to be overcome, and thus will also require higher temperature as well to drive the process forward. This is because initially there is a requirement of the reaction to merge protons and electrons to form hydrogen adatom on the surface.~\cite{tafel05} Thus the process will therefore either go through an associative and or dissociative Heyrovsky-type reaction. 

In Heyrovsky-type reactions~\cite{heyrovsky27}, the adsorbed species are known to be directly protonated in order to form a coordinated bond between the proton and the species. These protons directly attach to molecules from the electrolyte and then electrons merge from the surface with the protons to form a hydrogen bonded to the molecule. By applying a bias in later steps of the mechanism, a thermochemical barrier can be directly affected efficiently. Ideally, this study considers the possibility of reactions to take an associative Heyrovsky mechanism which proves to be more efficient from a biological standpoint than that of the dissociated mechanism. The associative Heyrovsky mechanism is similar to that of the mechanism for BNF, where the N-N bond is initially weakened by successive protonations until the dissociation barrier is low enough so that the N-N bond can be broken latte in the reaction steps. For this Heyrovsky mechanism, a nitrogen molecule is first attached to the surface and is then protonated before N-N bond dissociates. The below equations express reaction steps of BNF, which is the associative Heyrovsky mechanism (asterisk '*', denotes attachment to the surface),

\begin{equation}
N_{2}(g)+6(H^{+}+e^{-})+* \rightleftharpoons N_{2}*+6(H^{+}+e^{-}),
\label{equ:4}
\end{equation}
\begin{equation}
N_{2}*+6(H^{+}+e^{-}) \rightleftharpoons N_{2}H*+5(H^{+}+e^{-}),
\label{equ:5}
\end{equation}
\begin{equation}
N_{2}H*+5(H^{+}+e^{-}) \rightleftharpoons N_{2}H_{2}*+4(H^{+}+e^{-}),
\label{equ:6}
\end{equation}
\begin{equation}
N_{2}H_{2}*+4(H^{+}+e^{-}) \rightleftharpoons N_{2}H_{3}*+3(H^{+}+e^{-}),
\label{equ:7}
\end{equation}
\begin{equation}
N_{2}H_{3}*+3(H^{+}+e^{-})+* \rightleftharpoons 2NH_{2}*+2(H^{+}+e^{-}),
\label{equ:8}
\end{equation}
\begin{equation}
2NH_{2}*+2(H^{+}+e^{-}) \rightleftharpoons NH_{3}*+NH_{2}*+(H^{+}+e^{-}),
\label{equ:9}
\end{equation}
\begin{equation}
NH_{3}*+NH_{2}*+(H^{+}+e^{-}) \rightleftharpoons 2NH_{3}*,
\label{equ:10}
\end{equation}
\begin{equation}
2NH_{3}*\rightleftharpoons NH_{3}*+NH_{3}(g)+*,
\label{equ:11}
\end{equation}
\begin{equation}
NH_{3}*+NH_{3}(g)\rightleftharpoons 2NH_{3}(g)+*.
\label{equ:12}
\end{equation}
Addition of the fourth H to the N$_{2}$H$_{3}$* molecule allows a weakening of the N-N bond to readily dissociate molecules into the NH$_{x}$ species on the surface. In addition, there is a possibility of reaction (\ref{equ:7}) to split into NH and NH$_{2}$ on the surface and has been observed on some metals.\cite{howalt13}

The second mechanism, dissociated Heyrovsky mechanism, is also considered and compared to that of the associative mechanism. Where in this type of mechanism the nitrogen molecule is initially dissociated on the surface and then followed by subsequent protonation by direct attachment of protons. Below equations express reaction steps of Haber-Bosch process, which is the dissociated Heyrovsky mechanism,
\begin{equation}
N_{2}(g)+6(H^{+}+e^{-})+* \rightleftharpoons N_{2}*+6(H^{+}+e^{-}),
\label{equ:13}
\end{equation}
\begin{equation}
N_{2}*+6(H^{+}+e^{-})+* \rightleftharpoons 2N*+6(H^{+}+e^{-}),
\label{equ:14}
\end{equation}
\begin{equation}
2N*+6(H^{+}+e^{-}) \rightleftharpoons NH*+N*+5(H^{+}+e^{-}),
\label{equ:15}
\end{equation}
\begin{equation}
NH*+N*+5(H^{+}+e^{-}) \rightleftharpoons NH_{2}*+N*+4(H^{+}+e^{-}),
\label{equ:16}
\end{equation}
\begin{equation}
NH_{2}*+N*+4(H^{+}+e^{-}) \rightleftharpoons NH_{3}*+N*+3(H^{+}+e^{-}),
\label{equ:17}
\end{equation}
\begin{equation}
NH_{3}*+N*+3(H^{+}+e^{-}) \rightleftharpoons NH_{3}*+NH*+2(H^{+}+e^{-}),
\label{equ:18}
\end{equation}
\begin{equation}
NH_{3}*+NH*+2(H^{+}+e^{-}) \rightleftharpoons NH_{3}*+NH_{2}*+(H^{+}+e^{-}),
\label{equ:19}
\end{equation}
\begin{equation}
NH_{3}*+NH_{2}*+(H^{+}+e^{-}) \rightleftharpoons 2NH_{3}*,
\label{equ:20}
\end{equation}
\begin{equation}
2NH_{3}* \rightleftharpoons NH_{3}*+NH_{3}(g)+*,
\label{equ:21}
\end{equation}
\begin{equation}
NH_{3}(g)+NH_{3}* \rightleftharpoons 2NH_{3}(g)+*.
\label{equ:22}
\end{equation}

\subsection{Electrochemical Reaction}
DFT calculations approximations of the Gibbs free energy for the adsorbed species relative to the gas phase molecules of nitrogen and hydrogen, which can be obtained from the expression,
\begin{equation}
\Delta G = \Delta E + \Delta E_{ZPE} - T\Delta S,
\label{equ:22a}
\end{equation}
where $\Delta$E is the DFT calculated enthalpy, $\Delta$E$_{ZPE}$ is the reaction zero point energy, and $\Delta$S is the reaction entropy. In this study, only the ZPE is considered for the gas phases.

The driving electrochemical reaction is set to that of the standard hydrogen electrode (SHE) for the applied reference potential, which this study also takes into account in addition to zero point energy. This study is able to include the effect of the potential on the expressed reactions for surface sites by using the computational standard hydrogen electrode.

Thus the reference potential used by this study is that of the standard hydrogen electrode, which expresses the Gibbs free energy per hydrogen (chemical potential of (H$^{+}$+e$^{-}$) as related to that of $\frac{1}{2}$H$_{2}$(g), which is Equation \ref{equ:2} in the state of equilibrium. This implies that the pH = 0, the potential is that of U = 0 V relative to the SHE, and a pressure of 1 bar of H$_{2}$ in gas phase at 298 K, thus reaction (\ref{equ:1}) Gibbs free energy is equal to that of net reactions of (\ref{equ:4})-(\ref{equ:12}) and or (\ref{equ:13})-(\ref{equ:22}) at an electrode. 

\subsection{Computational Details}
Thermodynamic ground state properties were approximated for thermodynamic steps that assisted this study to analyze various configuration reaction steps by means of density functional theory (DFT) approach.~\cite{qe}DFT calculations relied on functional form of the pseudo-wave function that is based on Perdew-Burke-Ernzerhof (PBE) exchange-correlation function at potentials with a cut-off wave function energy of 1.5 keV (110 Ry) and a density cut-off radius of 1.8 keV (1320 Ry), which proved to be very accurate and stable for the studied unit cells. In addition, an ultra-soft pseudopotential and pseudized wave function were implemented to allow the reduction of computational expense for which this study greatly aided in the overall wall time required as a result of the large number of configurational combinations. Each unit cell was initially relaxed in the bulk structure and then subsequently setup as a slab and relaxed again. A Monkhorst-Pack with a k-point mesh sampling 2x2x2 grid with an offset of 1/2,1/2,1/2 was used for the bulk structure initial relaxation. A gamma k-point was used for the slab geometries with a super cell size of 2x2x1. A 10 {\r{A}} vacuum was placed on top of the slab. The slab geometry was relaxed prior to the introduction of the adsorbates. 

By solving the electronic densities self-consistently, adsorbates (N, H, NH, etc.) situated above the structure were relaxed to a relative total energy less than 1x10$^{-10}$. The unit cell was relaxed with the vacuum before any adsorbates were implemented in the structure. Then once relaxation of the structure in the vacuum was done, the substrate atoms were kept fixed and the adsorbates were allowed to relax on the surface of the structure in the 10 {\r{A}} vacuum. The reader should note that DFT predictions of energies, band gap. are often under predicted due to the over-analyticity of the functionals and exchange-correlation terms that need to be calculated. Thus the reported energies in this paper should not be used as absolute but used to study the trends for various material configurations.

\subsection{Experimental Synthesis}
\begin{figure}[ht]
\includegraphics[width=1.0\columnwidth,trim=3.5cm 0cm 2.6cm 1cm, clip=true]{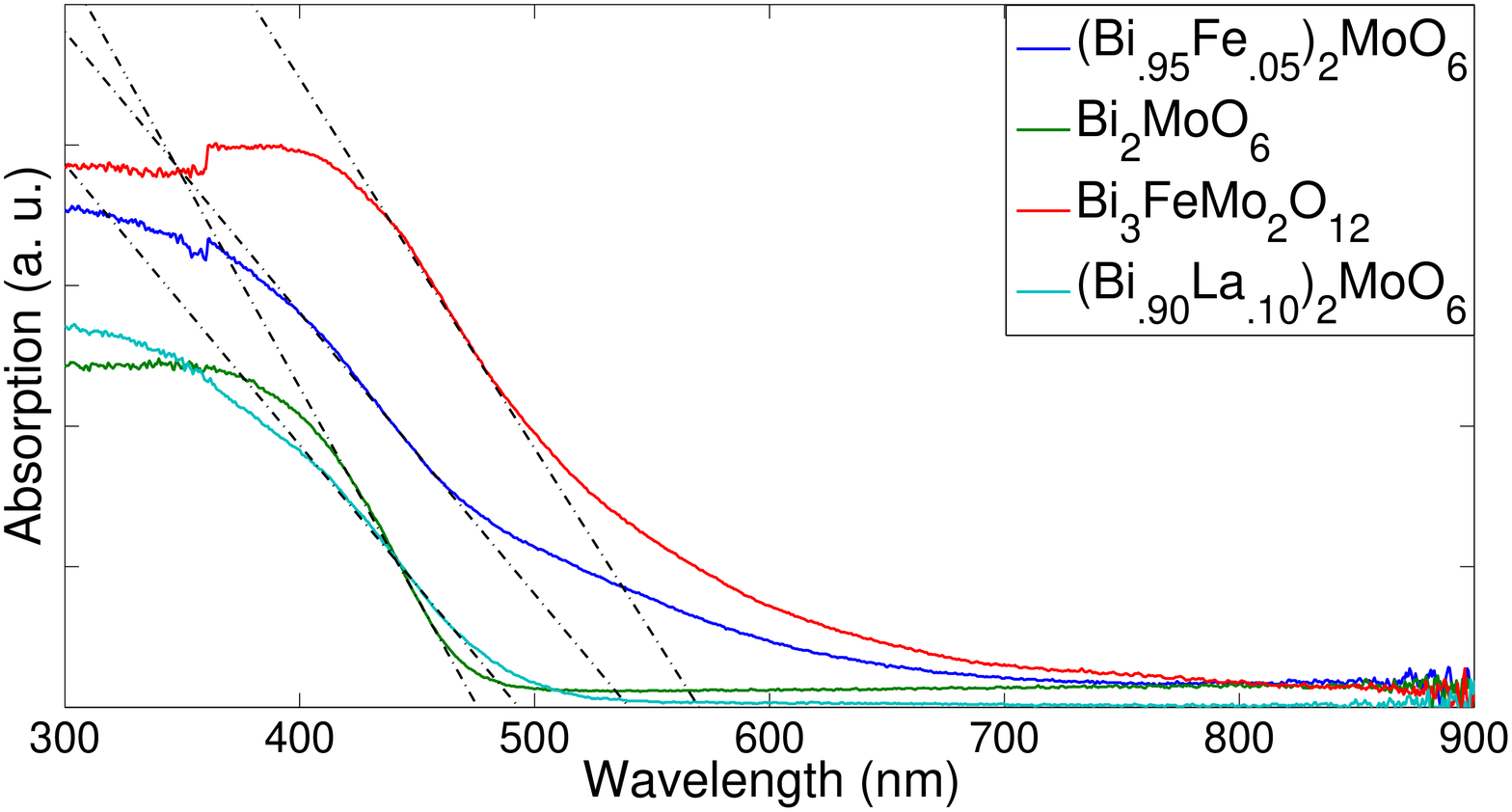}
\caption{Plot of the experimental UV-Vis spectrum for orthorhombic structure of: Bi$_2$MoO$_6$, Bi$_2$MoO$_6$ doped with 10\% La, Bi$_2$MoO$_{6}$ doped with 5\% Fe, and monoclinic structure of: Bi$_3$FeMo$_2$O$_{12}$. The fraction of (Bi$_{x}$M$_{y}$)$_2$MoO$_6$ (M = Fe, La, Yb) correspond to the percentage of Fe, La, Yb implemented in structure synthesis for the eight positions available for substitution. The corresponding experimental band gap values can be found in Table \ref{table:band_gap}.}\hrule
\label{fig:uv_vis}
\end{figure}

Materials synthesis based on hydrothermal method, and are used to express the calculated results for (Bi$_{x}$M$_{y}$)$_2$MoO$_6$ in Figure~\ref{fig:xrd} and Figure~\ref{fig:uv_vis}, and follows the listed procedures. Firstly for orthorhombic Bi$_2$MoO$_{6}$ structure: Na$_2$MoO$_4$*2H$_2$O was dissolved in 60 mL DI-water. Then Bi(NO$_3$)$_3$*5H$_2$O was slowly added to the above solution under vigorous stirring will pH was adjusted to 4-6. After stirring about 15 min, the suspension was transferred into 80 mL autoclaves and keep at 180 °C for 20 h. Then when cooling down to room temperature, the product was centrifuged and washed three times with DI-water and ethanol. Finally, the powder was obtained by drying the solution at 60 °C for 12 hours. For the 10\% Lanthanum doped Bi$_2$MoO$_{6}$ and the 5\% Iron doped Bi$_2$MoO$_{6}$ orthorhombic structure, an additional step was taken to add 10\% La or 5\% Fe respectfully into the precursor with same procedures taken as expressed above. Exactly the same steps are taken for Bi$_3$FeMo$_2$O$_{12}$ monoclinic structure synthesis with the addition of 25\% Fe into the precursor solution.

Shimadzu 2550 UV−Visible spectrometer equipped with an integrating sphere (UV 2401/2, Shimadzu) was used to obtain the absorption spectra under the diffuse reflection mode as shown for select compositions in Figure~\ref{fig:uv_vis}. XRD patterns are shown in Figure~\ref{fig:xrd} were recorded by X-ray diffraction (XRD, X′ Pert Pro PW3040-Pro, Panalytical Inc.) with Cu Kα radiation.

\begin{figure}[!ht]
\includegraphics[width=1.0\columnwidth,trim=6.5cm 0cm 7.6cm 1cm, clip=true]{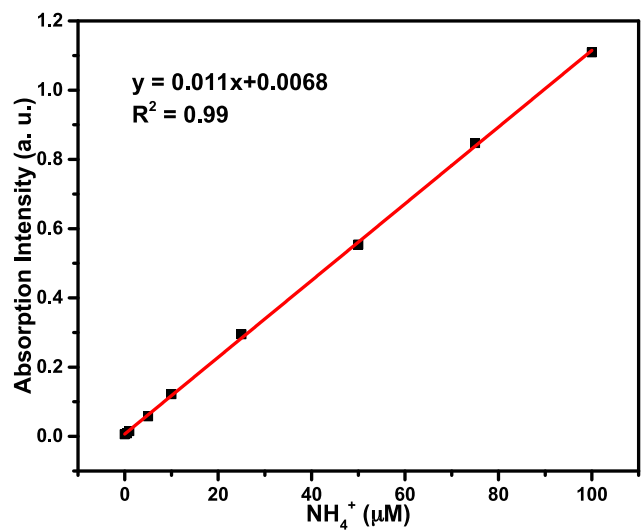}
\caption{Calibration curve of ammonium by Indophenol blue method from standard ammonium chloride solution. Also known as the standard curve, which has a linear relationship between absorption intensity and standard ammonium concentration. Used to determine concentration of a substance in an unknown sample by comparing the unknown concentration of a substance to a standard sample of known ammonium concentration.}\hrule
\label{fig:cal_curve}
\end{figure}

The photocatalysis experiment is as follows, 50 mg catalysts disperse in 100 mL DI-water. Using high pressure Hg lamp as light source. N$_2$ was bobbled 30 min before illuminated with light. During the photocatalytic reaction, N$_2$ was bobbling with 50 mL/min flow rate. 3 mL solution was taken out every 15 min. The concentration of ammonia was quantified by Indophenol blue method. The calibration curve is as follows, standard ammonium solution, phenol ethanol solution, sodium nitroprusside solution, sodium hypochlorite, mixture of sodium hydroxide and sodium citrate solution were pre-prepared. Oxidation solution was made by mix sodium hypochlorite and sodium citrate, sodium hydroxide solution before use.  Standard ammonium solution was diluted to 0.5 $\mu$M, 1 $\mu$M, 5 $\mu$M, 10 $\mu$M, 25 $\mu$M, 50 $\mu$M, 75 $\mu$M and 100 $\mu$M respectively. 3 mL standard solution mixed with phenol, sodium nitroprusside and oxidation solution. Then maintained in the dark at room temperature for at least 1 hour to generate color. Then used DI-water as the reference and attained absorption intensity from spectrometer. 

\begin{figure}[!ht]
\includegraphics[width=1.0\columnwidth,trim=6.5cm 0cm 7.6cm 1cm, clip=true]{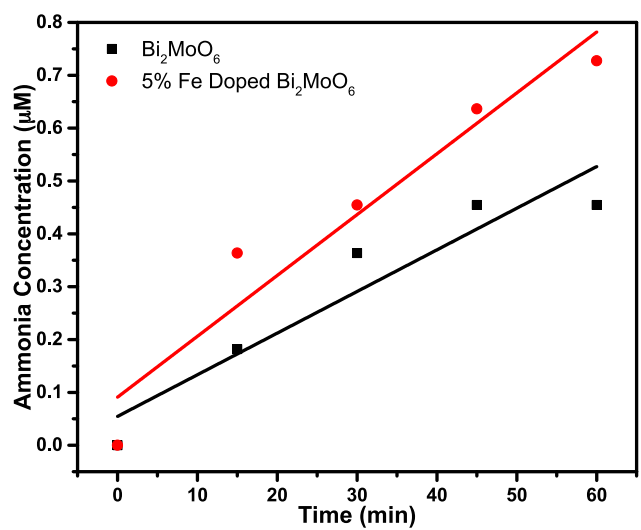}
\caption{Plot of ammonia concentration (production rate) vs. time maintained at room temperature to generate color for ammonia quantification of Bi$_2$MoO$_6$ and 5\% Fe doped Bi$_2$MoO$_6$. Catalysts concentration 0.5 g/L, ammonia rate is 4.7 $\mu$Mh-1 for Bi$_2$MoO$_6$ and 6.9 $\mu$Mh-1 for 5\% Fe doped Bi$_2$MoO$_6$.}\hrule
\label{fig:quantif}
\end{figure}

Based on linear relationship between absorption intensity and standard ammonium concentration, a calibration curve is developed for quantification of ammonia in the reaction, shown in Figure~\ref{fig:cal_curve}. From Figure~\ref{fig:quantif}, for bare Bi$_2$MoO$_6$, ammonia product rate is 4.7 $\mu$Mh-1 (0.5 g/L catalysts). Compared with bare sample, 5\% Fe doped Bi$_2$MoO$_6$ has ammonia product rate, which is 6.9 $\mu$Mh-1 (0.5 g/L catalysts) higher than that of undoped one. This experimental results illustrates that the dopant Fe could enhance N$_2$ activation and reduction, which is consistent with our computational results as expressed in the result sections.

\section{Results and Discussion}
All reaction states of the adsorption energies in (\ref{equ:4})-(\ref{equ:22}) are calculated for the various combinations of (Bi$_{x}$M$_{y}$)$_2$MoO$_6$ (M = Fe, La, Yb) in the orthorhombic structure and the monoclinic structure. These results are used to estimate Gibbs free energy change in elementary reactions (\ref{equ:4})-(\ref{equ:12}) for the associative mechanism and (\ref{equ:13})-(\ref{equ:22}) for the dissociative mechanism. Gibbs free energy differences of elementary steps assisted in approximating the theoretical over-potential needed to overcome for the reaction to evolve.

\subsection{Adsorption Sites}
\begin{figure}[h]
\includegraphics[width=1\columnwidth]{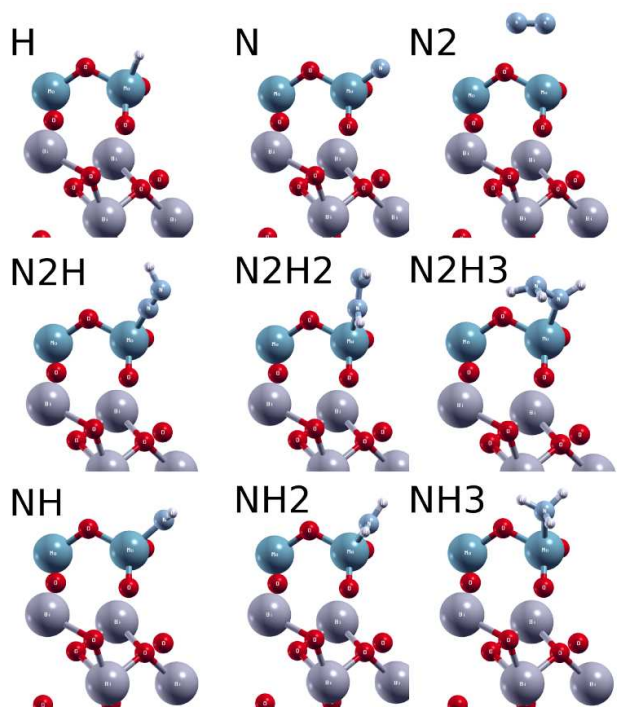} 
\caption{Adsorption sites of the orthorhombic Bi$_2$MoO$_6$ structure configuration in the [010]. The expressed adsorption sites for the Bi$_2$MoO$_6$ orthorhombic structure is only used to demonstrate the adsorption types and not for actual bonding for all other metals in this study. This means that the species of N$_{x}$H$_{y}$ could bind more strongly to Fe, La, Yb orthorhombic and monoclinic structure configurations. The corresponding color for atoms are red = O, white = H, dark blue = Mo, light blue = N, and gray = Bi.}\hrule
\label{fig:ads}
\end{figure}

Figure~\ref{fig:ads} illustrates the adsorption sites for the orthorhombic structure of Bi$_2$MoO$_6$, this is used to demonstrate characteristics of adsorption states for all configurations in the orthorhombic and monoclinic structures studied in this report. The orthorhombic and monoclinic structures demonstrate similar adsorption site trends, this is not to say that the binding energy is the same for different configurations of the orthorhombic and monoclinic structures. It is sufficient to say that combinations of (Bi$_{x}$M$_{y}$)$_2$MoO$_6$ (M = Fe, La, Yb) in the orthorhombic and monoclinic structures produce similar variations in adsorption sites. The expressed sites in Figure~\ref{fig:ads} are usually classified as adsorption sites of either bridge, hollow, and on top. It is also noted that typically each classification will yield slightly different geometries. Thus each structure of (Bi$_{x}$M$_{y}$)$_2$MoO$_6$ (M = Fe, La, Yb) in the orthorhombic and monoclinic configurations will yield unique electronic properties that can be analyzed and studied, which allows structures that prove more stable to be found and investigated.

Figure~\ref{fig:ads} illustrates the adsorption sites on Bi$_2$MoO$_6$ orthorhombic structure surface. The hydrogen molecule proved to be stable on the exposed surface layer of metal sites and tended to lie on the crystal surface directly attaching to one metal atom. Thus allowing hydrogen to have a bonding site of bridge classification, in fact a very small difference of energies was found between adsorption sites that responded to hydrogen bonding. Nitrogen adsorbed to the surface on a hollow site and preferred to bind in a sigma (strong covalent bond, formed by overlapping between atomic orbitals) bond on the surface with two metal atoms, thus proving edge sites to be the most stable with respect to all other sites. However, this was not true for adsorption of N$_{2}$, this is because N$_{2}$ molecule prove to have limited stable adsorption configurations, due to the strong N-N bond and proved to have a bonding site of top classification.

The NH molecule was found adsorbed on the surface in hollow sites similar to N, NH$_{2}$ to bridge sites, and NH$_{3}$ on-top sites and all proved to be stable structures (Figure~\ref{fig:ads} bottom row). However, in the case of N$_{2}$H, N$_{2}$H$_{2}$, and N$_{2}$H$_{3}$ species (Figure~\ref{fig:ads} middle row), they typically preferred to bond in a bridge site, and only tend to favor the surface as more hydrogen was introduced in the species, one of the nitrogen atoms bonded to a metal atom similarly to the sigma bonding expressed for N and NH$_{x}$ species (Figure~\ref{fig:ads} top and bottom row). With respect to all the adsorption sites, species' configuration orientation changed slightly as more hydrogen atoms are implemented in the surface species. When looking carefully at Figure~\ref{fig:ads}, weakening of the N-N bond is demonstrated as a visual representation of the internal bonding length dramatically increases as more hydrogen atoms are implemented in the species. This can be seen when looking at NH$_{3}$ adsorbed on the structure, where the internal bonding length is much greater to that of other species as a result of the higher ratio of hydrogen to nitrogen. Also, it is very apparent that nitrogen atom becomes closer to the metal atom as more hydrogen atoms become bonded to the respected nitrogen atom as seen in Figure~\ref{fig:ads} middle row, this is demonstrated for BNF pathway and shows the weakening of N-N bond.

\begin{figure*}[!ht]
\includegraphics[width=.99\columnwidth,trim=2.5cm .5cm 3cm 1cm, clip=true]{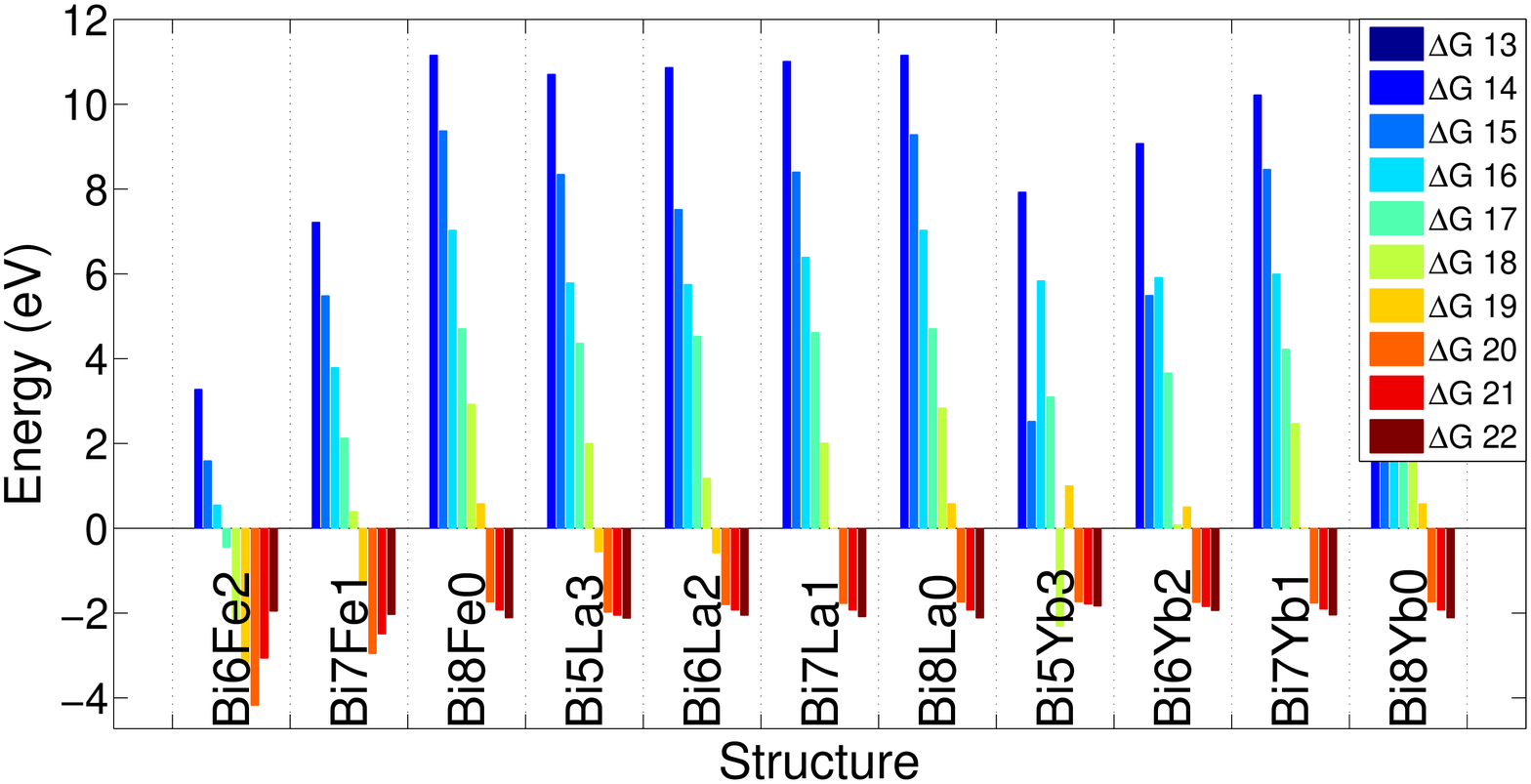}
\includegraphics[width=.99\columnwidth,trim=2.5cm .5cm 3cm 1cm, clip=true]{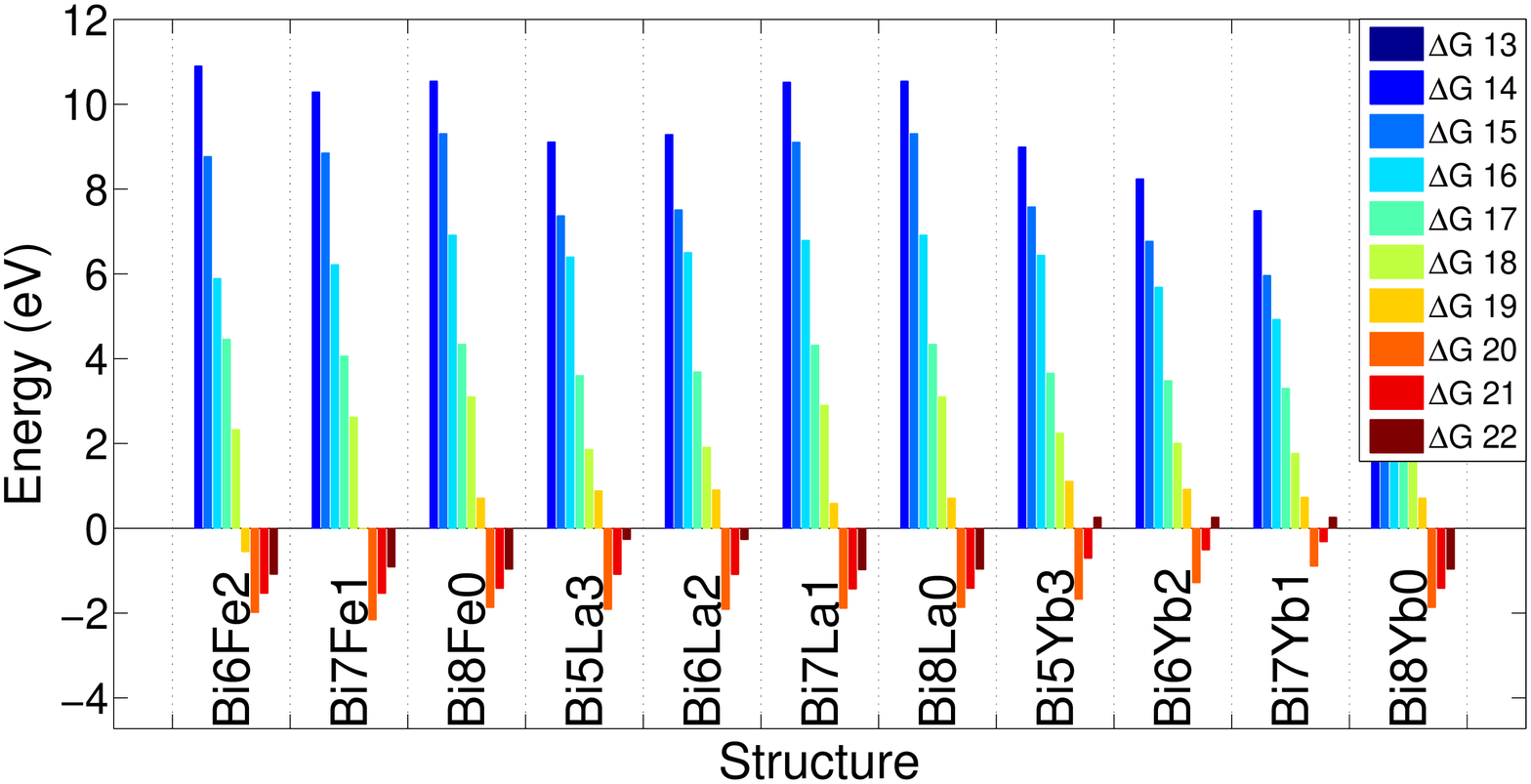}\\
\vspace{-1cm}
\begin{flushleft}
\textbf{\begin{Large}A\end{Large}}\hspace{87mm}\textbf{\begin{Large}C\end{Large}} 
\end{flushleft}
\includegraphics[width=.99\columnwidth,trim=2.5cm .5cm 3cm 1cm, clip=true]{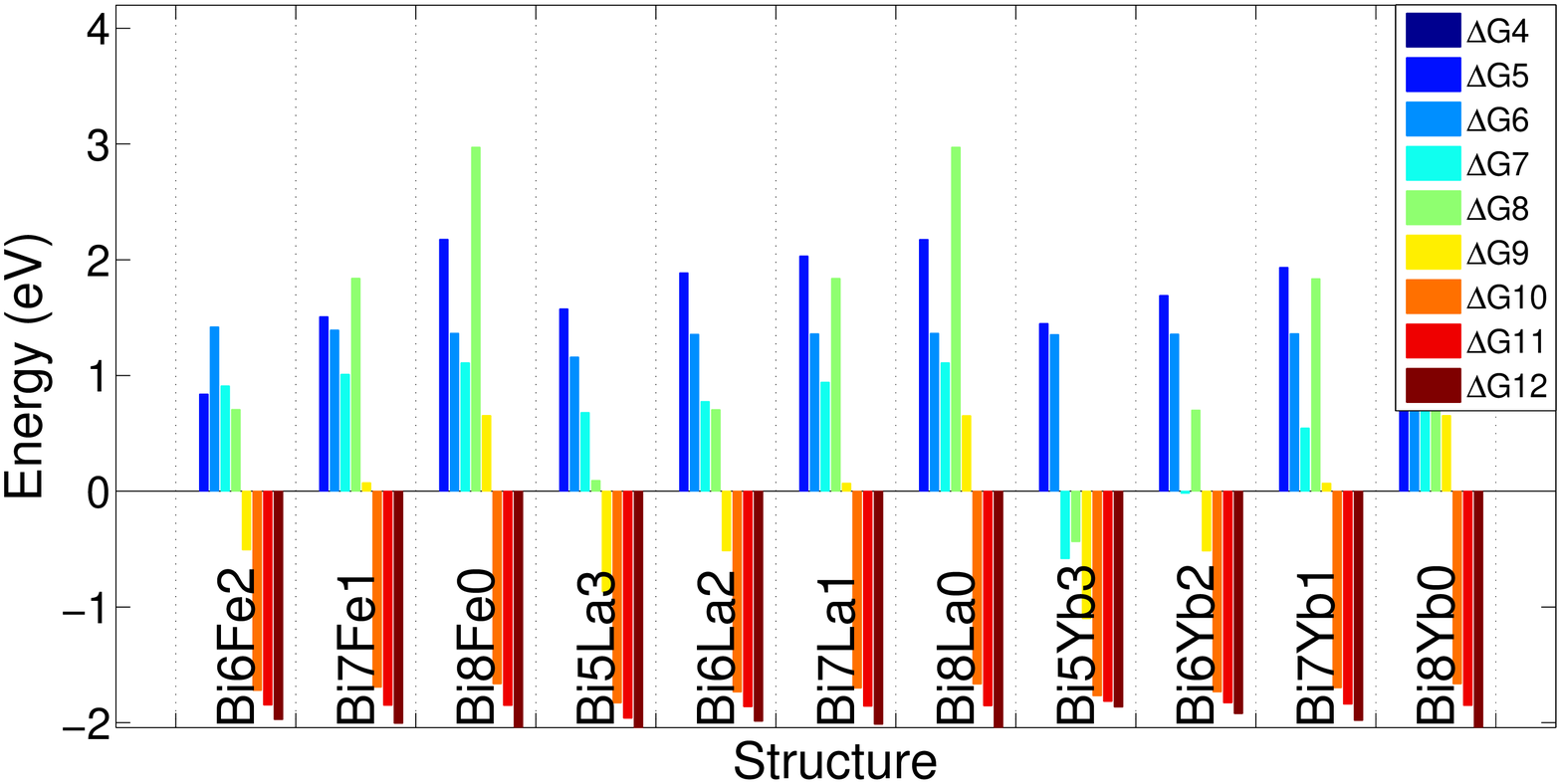}
\includegraphics[width=.99\columnwidth,trim=2.5cm .5cm 3cm 1cm, clip=true]{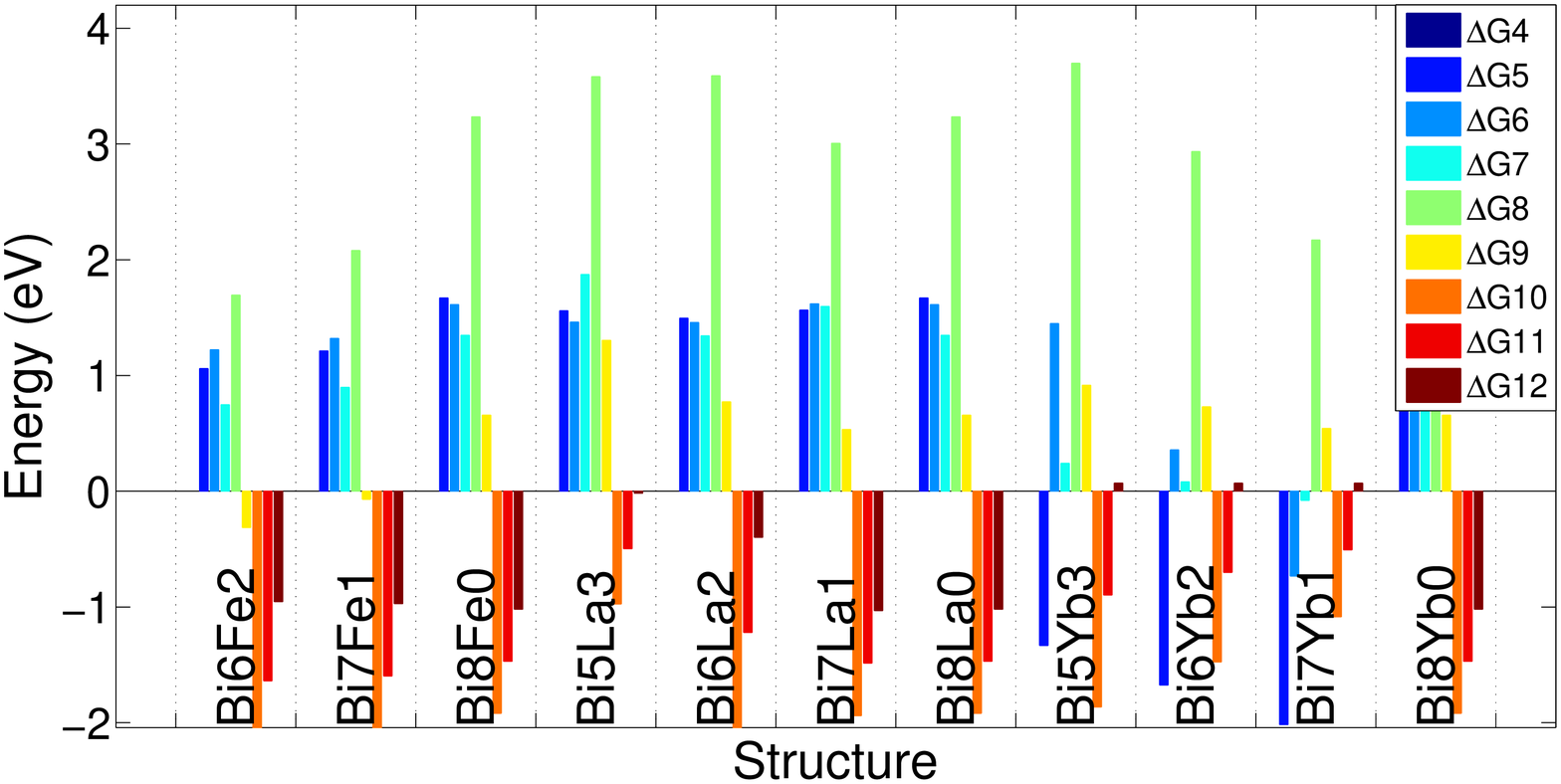} \\
\vspace{-1cm}
\begin{flushleft}
\textbf{\begin{Large}B\end{Large}}\hspace{87mm}\textbf{\begin{Large}D\end{Large}} 
\end{flushleft}
\caption{Cumulative Gibbs energy for the dissociative (A,C) and associative (B,D) mechanism on both the orthorhombic (A,B) and monoclinic (C,D) structure configurations. The respected energies are attained from DFT calculations for an electrolyte with pH=0 at T=0 K. The expressed energies correspond to relative reaction steps of (\ref{equ:4})-(\ref{equ:12}) for the associative mechanism and (\ref{equ:13})-(\ref{equ:22}) for the dissociative mechanism . Each configuration can be associated with the potential-energy curve for successive reaction steps. The best performing structure for the associative mechanism is the orthorhombic configuration of (Bi$_{6/8}$Fe$_{2/8}$)$_2$MoO$_6$ (B) and for the dissociative mechanism is the orthorhombic configuration of (Bi$_{6/8}$Fe$_{2/8}$)$_2$MoO$_6$ (A).}\hrule
\label{fig:step}
\end{figure*}

\subsection{Ammonia Formation on Surface}
Each surface composition of the orthorhombic and monoclinic configurations with the 9 adsorbate species of nitrogen and hydrogen that conformed to the two reaction pathways were investigated, which allowed this study to predict the electrochemical reaction of each step for the two mechanism pathways. Figure~\ref{fig:step} expresses the Gibbs free energy of the electrochemical reactions of steps (\ref{equ:4})-(\ref{equ:12}) (${\Delta}$G4-${\Delta}$G12) for both the associative and (\ref{equ:13})-(\ref{equ:22}) (${\Delta}$G13-${\Delta}$G22) the dissociative mechanism of orthorhombic and monoclinic structures in configurations of (Bi$_{x}$M$_{y}$)$_2$MoO$_6$ (M = Fe, La, Yb). It is noted that some combinations of the (Bi$_{x}$M$_{y}$)$_2$ MoO$_6$ in Figure~\ref{fig:step} proved to be unstable and propagated to much into the exposed vacuum after relaxation steps. Thus for the orthorhombic configurations, high Fe and Yb input into the structure is not present and for the monoclinic configurations, high Fe input into the structure is not present. This was also recorded in the experimental results for high Fe, Yb, and La input into both configurations, which could not be stabilized and lost the orthorhombic and monoclinic phase. Also, it should be noted that theoretical calculations are expressed for ideal conditions, were as experimental calculations are of practical conditions. Thus some of the theoretically stabilized configurations may prove to be less than ideal for synthesis, never the less this paper aims to express and compare calculations that better assist the explored structures.

Theoretical calculations express N$_{2}$ binding to almost all the surfaces of orthorhombic and some of the surfaces of monoclinic configurations of (Bi$_{x}$M$_{y}$)$_2$MoO$_6$ (M = Fe, La, Yb) structures, this resulted in adsorption energy that always expressed slightly negative reaction steps when compared to all other reaction steps as shown in Figure~\ref{fig:step}A,C dissociative ${\Delta}$G14 and Figure~\ref{fig:step}B,D associative ${\Delta}$G5 and ${\Delta}$G8. The larges loss in entropy is associated from steps leading to gas phase N$_{2}$ to surface bonded molecules, this large loss of energy corresponds to slightly negative Gibbs energy steps expressed for the orthorhombic and monoclinic structures, Figure~\ref{fig:step}. Most of the potential determining steps (largest positive step to overcome) in reducing N to form ammonia for both the associative and dissociative pathways in the structures, tended to be the breaking of the N-N bond as expressed in dissociative step ${\Delta}$G14 of Figure~\ref{fig:step}A,C for both structures and for the monoclinic associative step ${\Delta}$G8 of Figure~\ref{fig:step}D but not for the associative orthorhombic Figure~\ref{fig:step}B. This is because most of the monoclinic structures did not favor a direct attachment of N$_{2}$ on the surface, thus resulted in both the associative and dissociative pathways of the monoclinic structure to have the largest step requiring an external source to break the N-N bond. This is true for the dissociative mechanism where the N-N bond is broken early for the reaction thus the largest step is that of N-N bond braking as shown in Figure~\ref{fig:step}A,C ${\Delta}$G14. However, this is not true for the associative step of the orthorhombic configuration Figure~\ref{fig:step}B. Where the determining step is that of the addition of one hydrogen atom to transition from N$_{2}$ to N$_{2}$H, Figure~\ref{fig:step} from ${\Delta}$G5 to ${\Delta}$G6, because for the most part the N$_{2}$ tended to better attach on the surface in the orthorhombic configuration. The expressed hydrogenation step in the orthorhombic configuration is also expressed as a large energy step to overcome from the binding of H from its gas phase, which corresponds to the step taken to overcome Gibbs free energy formation of N$_{2}$ and NH$_{2}$ on the surface.

Looking at just the hydrogen steps, the first addition of hydrogen atom, from ${\Delta}$G5 to ${\Delta}$G6, typically is the most positive step in energy for most surfaces explored for all hydrogen steps. However, with respect to all steps taken for the dissociative mechanism of both configurations and associative mechanism of monoclinic configuration, the most significant over-potential is when initially breaking N-N bonding from N$_{2}$ to N, Figure~\ref{fig:step}A,C from ${\Delta}$G14 to ${\Delta}$G15 and Figure~\ref{fig:step}D from ${\Delta}$G7 to ${\Delta}$G8. This is expected due to the strong sigma bond expressed by overlapping atomic orbitals of N-N. For both configurations of both pathways the last three steps, Figure~\ref{fig:step}A,C going from ${\Delta}$G20 to ${\Delta}$G21 to ${\Delta}$G22 and Figure~\ref{fig:step}B,D going from ${\Delta}$G10 to ${\Delta}$G11 to ${\Delta}$G12 are all negative steps expressed by the reduction of NH$_{3}$ on the surface to NH$_{3(g)}$ from large gains in entropy associated with transitions of surface bounded molecule to gas phase.

\subsection{Adsorption of N$_{2}$H$_{x}$ and NH$_{x}$ Species on Fe, La, Yb Configurations}
Initially, the two reaction pathways rely on configurations that have adsorption steps of N$_{2}$H$_{x}$ molecules for the associative and NH$_{x}$ molecules for the dissociative mechanisms. Morphology and composition of the monoclinic and orthorhombic surface structures greatly contribute to the Gibbs energies for reactions steps taken of associative and dissociative mechanisms. Presented in this study are stable configurations of the monoclinic and orthorhombic bismuth containing surfaces as shown in Figure~\ref{fig:step} and Figure~\ref{fig:max_step}, which are noted to be stable theoretically by calculations of the formation energy and by direct experimental synthesis for some configurations of (Bi$_{x}$M$_{y}$)$_2$MoO$_6$ (M = Fe, La, Yb). Calculations predict that high concentrations of Fe, La, and Yb to prove unstable. The configurations that DFT calculation are performed on are for structures that have energies within the scale factor attained from experimental results.  The scale factor is a product of Boltzmann constant and temperature for the experimental sample (kT). Thus thermal energy greater than that of the formation energy for synthesized samples prove unstable. Also, experimental syntheses prove unstable for 40\% or more implementation of La, Yb, and Fe into the orthorhombic and monoclinic structures, and forms secondary phases.

\begin{figure}[!ht]
\includegraphics[width=1\columnwidth,trim=2.5cm 0cm 2.5cm 1cm, clip=true]{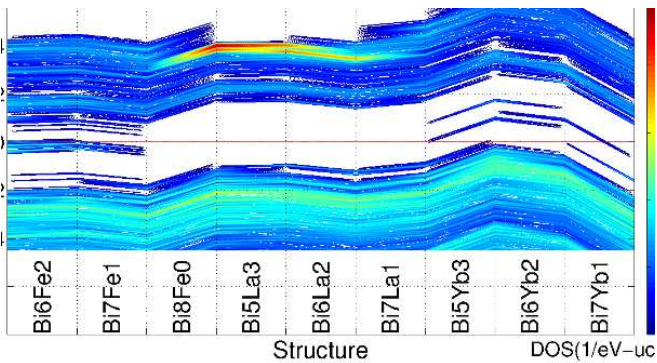}
\caption{Illustration of the density of states (DOS) calculated for the orthorhombic configurations of (Bi$_{x}$M$_{y}$)$_2$MoO$_6$ (M = Fe, La, Yb). DOS of monoclinic configurations proved to demonstrate similar trends when implementing Fe, La, and Yb, thus only orthorhombic configurations are shown. Note that all Fermi levels (red line) are aligned at 0 eV for direct comparison. Select band gap calculated values are expressed in Table~\ref{table:band_gap}. The trend suggests that more Fe and Yb implemented into the structure further reduces the band gap by introducing impurity states in the band gap.}\hrule
\label{fig:dos}
\end{figure}

For any La, Yb, and Fe implemented in both structures, that the determinant step (Figure~\ref{fig:max_step}) is associated with the breaking of the N-N bond as seen in Figure~\ref{fig:step}A,C of ${\Delta}$G14 for dissociative and D of ${\Delta}$8 for associative mechanisms, except for the associative mechanism of orthorhombic structure in Figure~\ref{fig:step}B. This is because the orthorhombic structure has a better affinity toward nitrogen attachment to the surface, due to the exposed surface of Mo sites tending to have much higher affinity for N species. For the orthorhombic structure of the associative mechanism, Figure~\ref{fig:step}B, the determinant step is associated with hydrogenation steps ${\Delta}$G4 to ${\Delta}$G5 which is the addition of the first H step to form N$_2$H on surface or ${\Delta}$G5 to ${\Delta}$G6 which is the addition of second H step form N$_2$H$_2$. This occurs because all configurations of the orthorhombic structure contain the same number of Mo atoms on the surface that directly interact with N species, Mo has higher affinity to N atom species and acts as a precursor to directly attach N species as seen in Figure~\ref{fig:ads} reference biological nitrogen fixation section for Mo affinity for N atom species.

\subsection{Onset Maximum Positive Determining Step}
Looking at the minimum positive potential for the orthorhombic and monoclinic structures one can directly observe that the most determinant step is associated with the breaking of N-N as expressed in Figure~\ref{fig:step}A,C,D. Which is expected for the dissociative mechanism as seen in Figure~\ref{fig:step}A,C where the N-N bond is broken initially. It's reasoned that the orthorhombic configurations behave relatively as expected for both mechanisms. However, when it comes to the monoclinic configurations of the associated mechanism, Figure~\ref{fig:step}D, the structure is still having trouble stabilizing N-N near the surface as indicated by the determining step associated with breaking of the N-N as seen for ${\Delta}$G8s of Figure~\ref{fig:step}D that correspond to all determinant step for that configuration.

For the orthorhombic structure, there are two scenarios that determine the best structure for nitrogen fixation, scenario (i) which configuration better stabilizes N atom on the surface and scenario (ii) how well can the structure attract H atom. Scenario (i) is needed for the initial breaking of N-N bond for the dissociative mechanism Figure~\ref{fig:step}A and scenario (ii) is needed for the successive initial protonation as seen in the associative mechanism Figure~\ref{fig:step}B. In both scenarios the answer lies with the Mo-O atoms in the orthorhombic structure and how well can the structure influence the Mo-O atoms to better assist the attraction of N and H species on the surface (Mo-O layer). The Mo-O layer is what the N and H species interact with the most as seen in Figure~\ref{fig:ads} because the location of the vacuum which was found to only stabilized in the [010] plane for orthorhombic structures (from exploring all possible orientations) as seen in Figure~\ref{fig:UC}A and thus forces that interaction of N species and H species with Mo-O layer.

\begin{figure*}[!ht]
\includegraphics[width=.99\columnwidth,trim=2.5cm .5cm 3cm 1cm, clip=true]{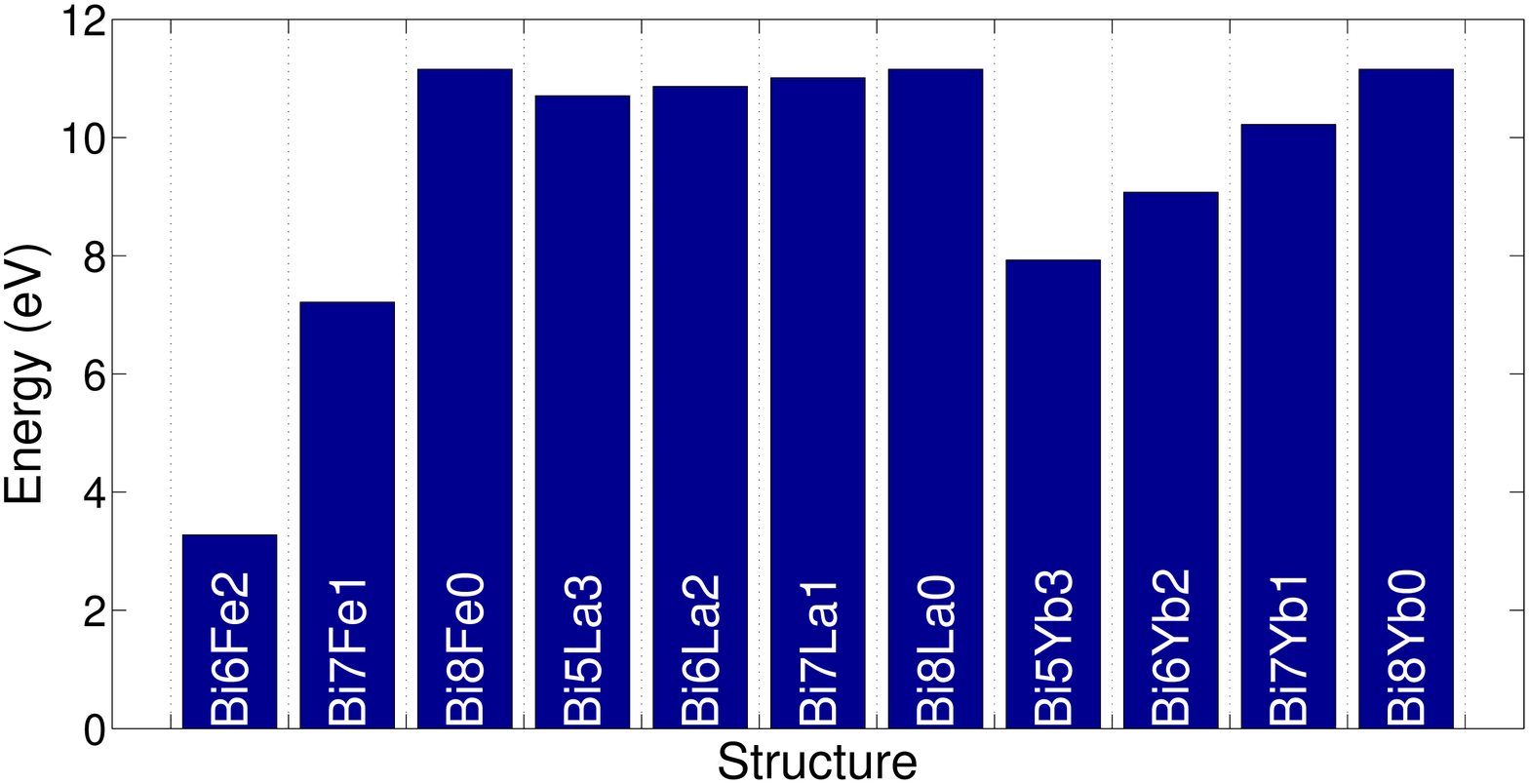}
\includegraphics[width=.99\columnwidth,trim=2.5cm .5cm 3cm 1cm, clip=true]{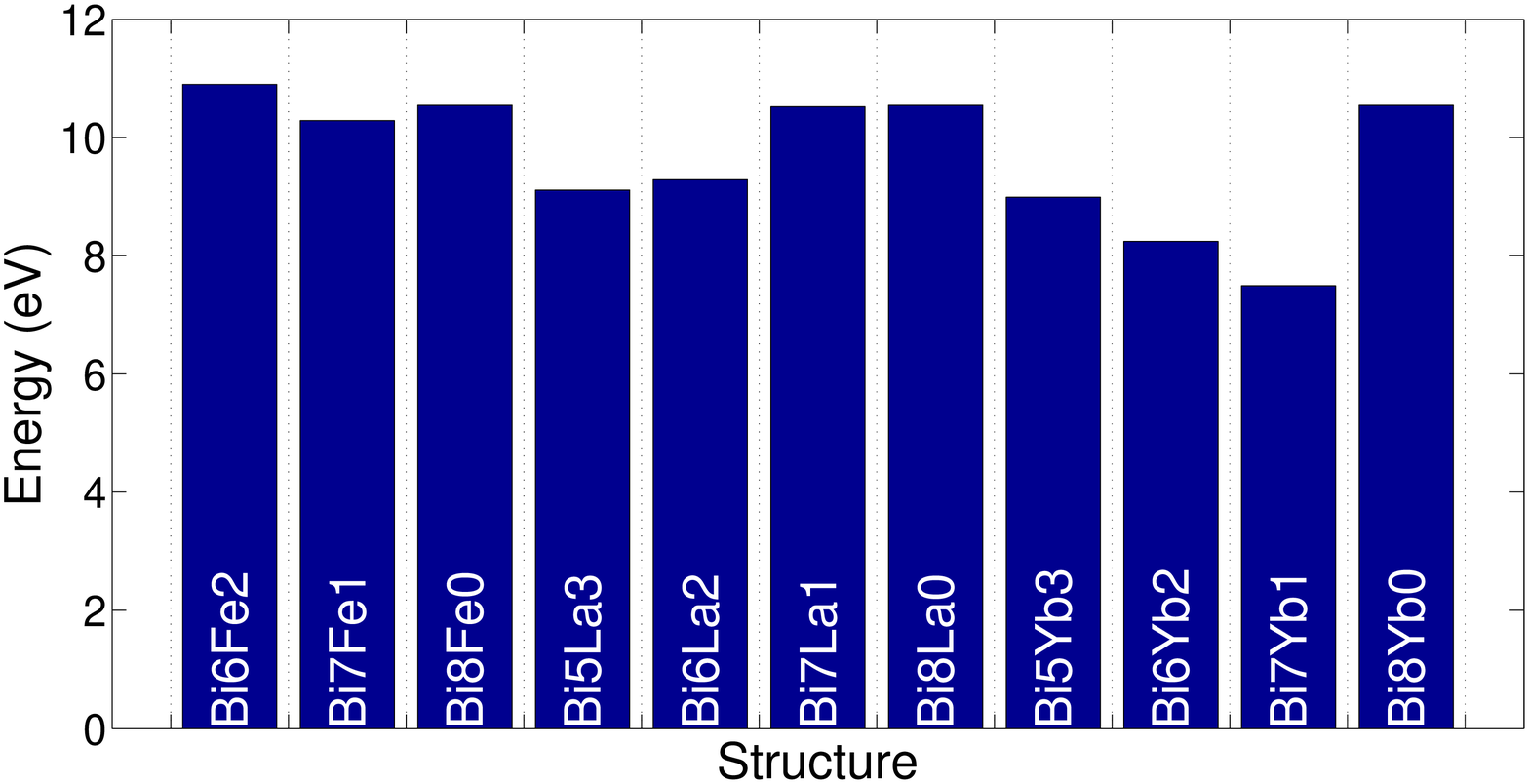} \\
\vspace{-1cm}
\begin{flushleft}
\textbf{\begin{Large}A\end{Large}}\hspace{87mm}\textbf{\begin{Large}C\end{Large}} 
\end{flushleft}
\includegraphics[width=.99\columnwidth,trim=2.3cm .5cm 3cm 1cm, clip=true]{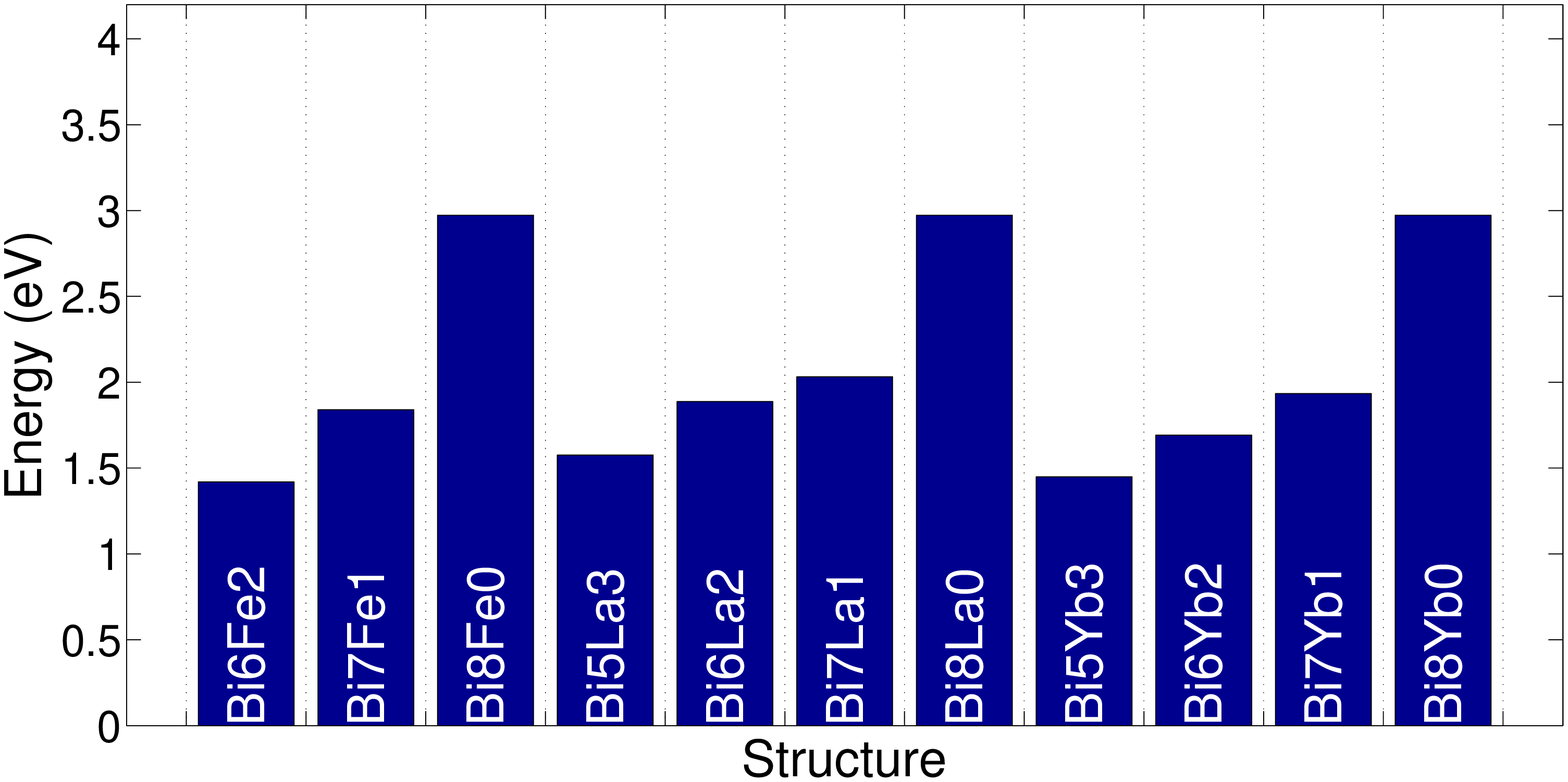}
\includegraphics[width=.99\columnwidth,trim=2.3cm .5cm 3cm 1cm, clip=true]{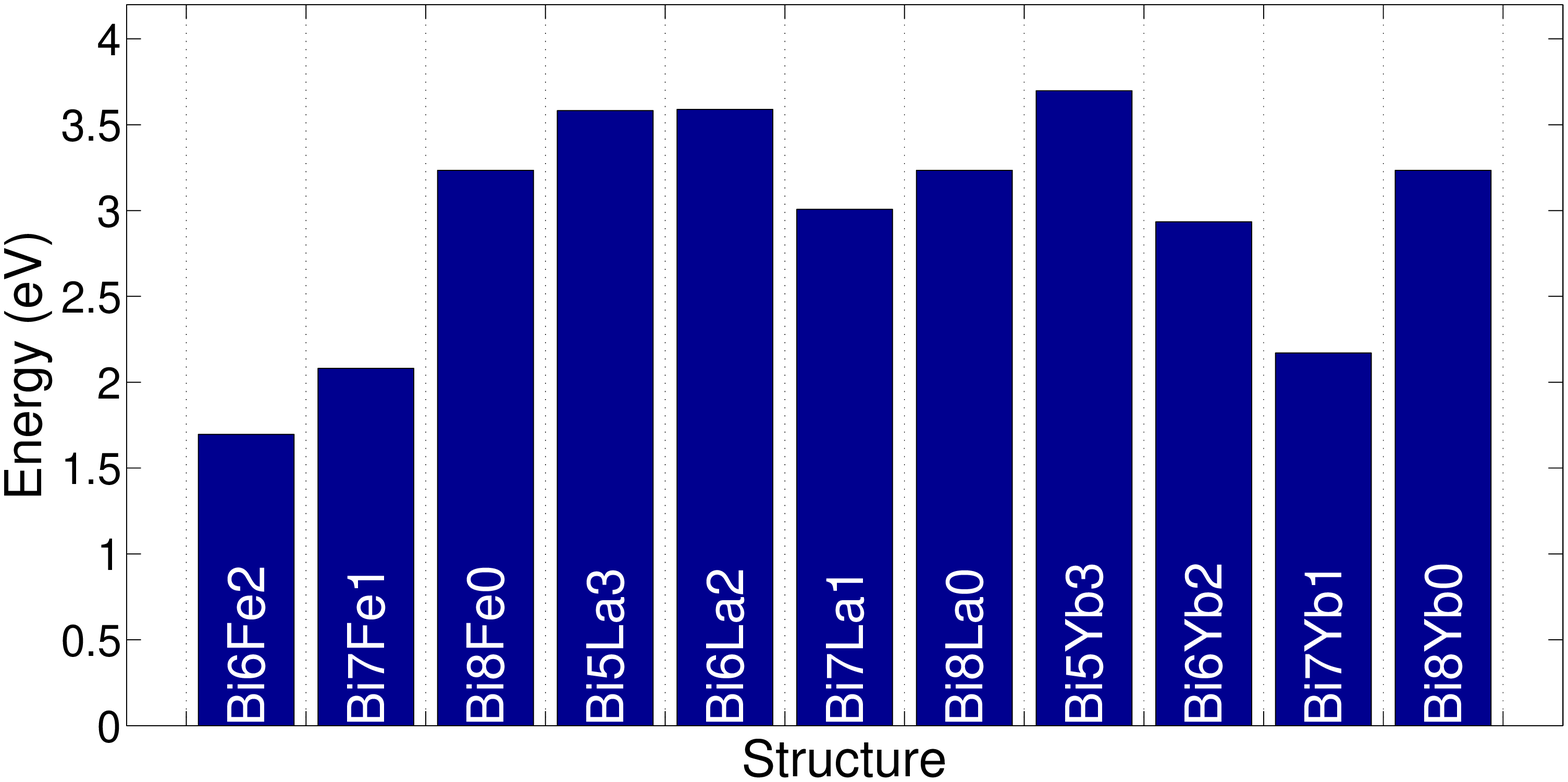} \\
\vspace{-1cm}
\begin{flushleft}
\textbf{\begin{Large}B\end{Large}}\hspace{87mm}\textbf{\begin{Large}D\end{Large}} 
\end{flushleft}
\caption{Plots of the maximum reaction barrier for all successive reactions for Fe, La, Yb implemented for the orthorhombic A,B and monoclinic C,D Bismuth-based structures. The bars corresponds to the energies of the most difficult step to overcome for reaction steps (\ref{equ:4})-(\ref{equ:12}) for the associative mechanism (B,D) and (\ref{equ:13})-(\ref{equ:22}) for the dissociative mechanism (A,C). The best composition for orthorhombic and monoclinic structures are tabulated in Table~\ref{table:best_points}.}\hrule
\label{fig:max_step}
\end{figure*}

\begin{table}
\small
\begin{center}
\begin{tabular}{ c |c |c }
\hline
 Design & DFT (eV) & Experimental UV-Vis (eV) \\
 \hline
 Bi$_2$MoO$_6$ 					&2.40 & 2.61\\
 (Bi$_{.95}$Fe$_{0.05}$)$_2$MoO$_6$ &2.11 & 2.30\\
 (Bi$_{.90}$La$_{0.10}$)$_2$MoO$_6$ &2.30 & 2.51\\
 Bi$_3$FeMo$_2$O$_{12}$ 			&1.80 & 2.17\\
 \hline
\end{tabular}
\end{center}
\caption{Calculated and experimentally determined band gap energies for Bismuth-based configurations of the orthorhombic and monoclinic structure. The DFT calculations attained from DOS Figure~\ref{fig:dos} and the experimental data was determined using UV-Vis spectrometer, see Figure~\ref{fig:uv_vis}. As expected of the DFT calculations, the overall band gap is under approximated but does demonstrate the correct trend as the experimental results.}
\label{table:band_gap}
\end{table}

\begin{figure}[!h]
\begin{center}
\includegraphics[width=.7\columnwidth]{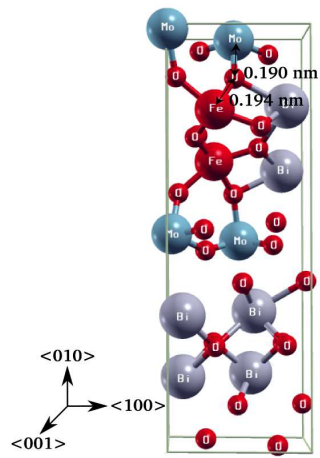} 
\end{center}
\caption{Illustration of the (Bi$_{6/8}$Fe$_{2/8}$)$_2$MoO$_6$ in orthorhombic configuration. Fe atoms induce localized distortions of Bi, Mo, O as shown, and this modulates the affinity of N and H species toward Mo atoms.}\hrule
\label{fig:FeBi6}
\end{figure}

Figure~\ref{fig:dos} better illustrates what happens to the orthorhombic configuration as more Yb, La, and Fe are added in the structure. It is quite apparent that more states are able to be introduced in the band gap as more Yb and Fe are implemented thus resulting in impurity states in the band gap. These impurity states are associated with strong orbital interactions due to strong hybridization between the Bi 6p and Fe 3d orbitals and also the O 2p and Fe 3d orbitals. This is due to the energetic overlap in the Fe 3d bands with the Bi 6p bands and the O 2p and Fe 3d bands, this result is also expressed in literature~\cite{oikawa05}. Adding Fe will induce localized distortions of the Mo atoms and this modulates the affinity of N and H species toward the Mo atoms. This can be seen in Figure~\ref{fig:FeBi6} for a Fe containing orthorhombic structure where enhancement of the bands due to hybridization causes strong interaction between Fe-O. The strong interaction can be visually represented with the decreased internal bonding length when comparing the same location of Fe-O of Figure~\ref{fig:FeBi6} with Bi-O in Figure~\ref{fig:UC}A which alters the bond length of Mo-O. This opens bonds to N and H species as shown in Figure~\ref{fig:UC}A and Figure~\ref{fig:FeBi6}, where Mo-O distance increases from 0.186 nm to 0.190 nm as Fe is implemented in the structure thus resulting in Mo interacting better with N and H species on the surface. This can also be said for the Yb 4f orbitals which demonstrate strong interaction with O and Bi thus also causing this shift in states. Meaning that all surrounding atoms like Bi, O, and Mo will be affected due to the substitution of Fe and Yb atoms. Thus resulting in a shifting of stats that eventually will cause the strong interaction of Mo-O layer to N and H species.

Fe and Yb in the lattice cause a shifting of states due to changes in inter-atomic bond lengths, which indirectly affect the interaction of N species to the surface of orthorhombic configurations that better assist the dissociative mechanism of Figure~\ref{fig:step}A. This indirect interaction expresses the best determinant step as (Bi$_{6/8}$Fe$_{2/8}$)$_2$ MoO$_6$ for all studied structures in this paper for the dissociative mechanism of approximately 3.2 eV as seen in Figure~\ref{fig:max_step}A, which is associated with N-N breaking ${\Delta}$G14. Any Fe implemented in the orthorhombic structure resulted in greatly reducing the band gap as seen in Table~\ref{table:band_gap} of the recorded experimental and DFT calculated band gap. The small band gap results in larger population of excited carriers as a result of increased utilization of the electromagnetic spectrum. The addition of Fe and Yb into the structure also results in changing the majority carrier concentration. This can be seen in Figure~\ref{fig:dos} as a shift in the Fermi level towards the valence band (HOMO) and conduction band (LUMO).  The same idea of the Mo layer strong effect can also be made for the orthorhombic configuration of the associative mechanism for H species, Figure~\ref{fig:step}B. The better structure for the orthorhombic associative pathway proved to be Bi4Fe2 of Figure~\ref{fig:dos} ((Bi$_{6/8}$Fe$_{2/8}$)$_2$MoO$_6$), which tended to stabilize N species on the surface and still maintain good affinity with the H species. Thus resulting in better protonation steps taken for all studied structures in the associative mechanism. The (Bi$_{6/8}$Fe$_{2/8}$)$_2$MoO$_6$ of orthorhombic configuration in Figure~\ref{fig:step}B had approximately 1.4 eV for the determinant step, which is the lowest for all studied structures in this paper that are associated with ${\Delta}$G6 of H atom attachment to N$_2$. 

\section{Partial Density of States}

\begin{figure}[!ht]
\includegraphics[width=.98\columnwidth]{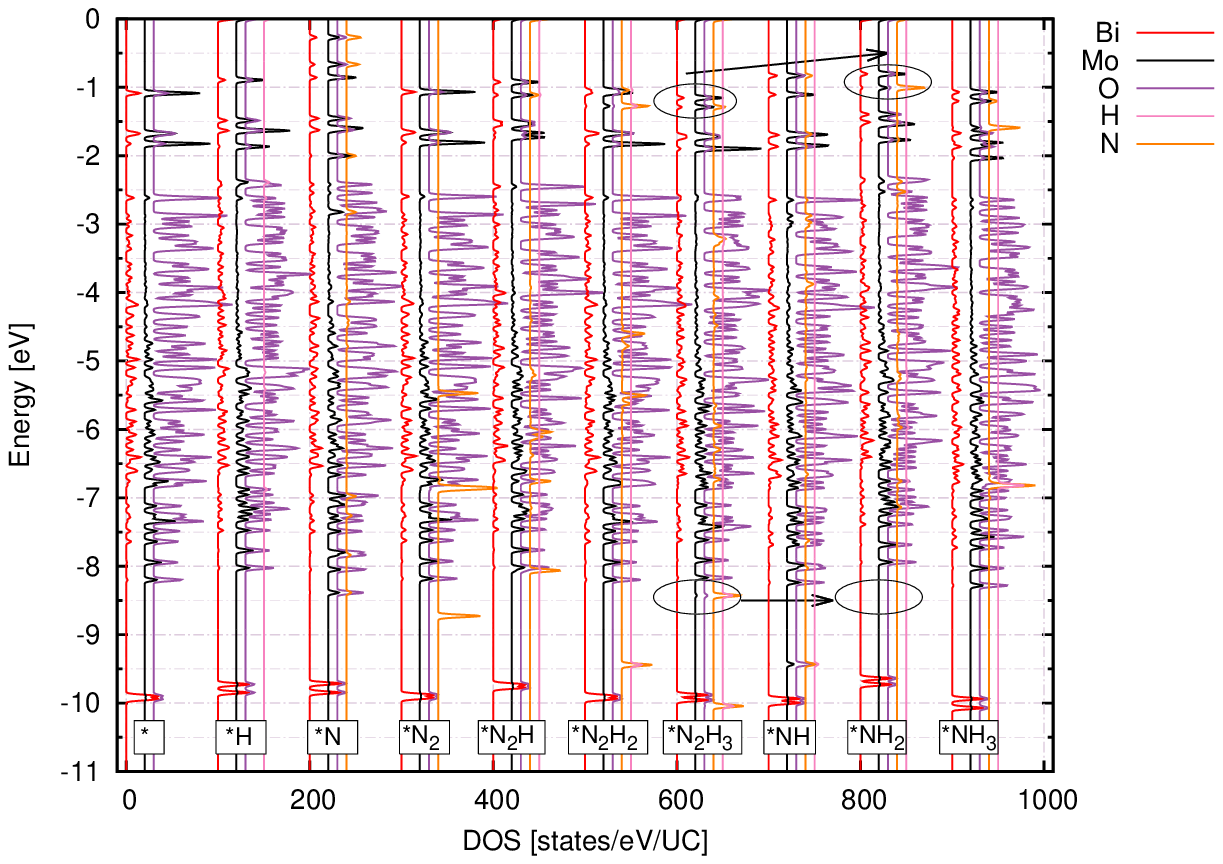} \\
\vspace{-1.8cm}
\begin{flushleft}
\textbf{\begin{Large}A\end{Large}} \\
\end{flushleft}
\vspace{.25cm}
\includegraphics[width=.98\columnwidth]{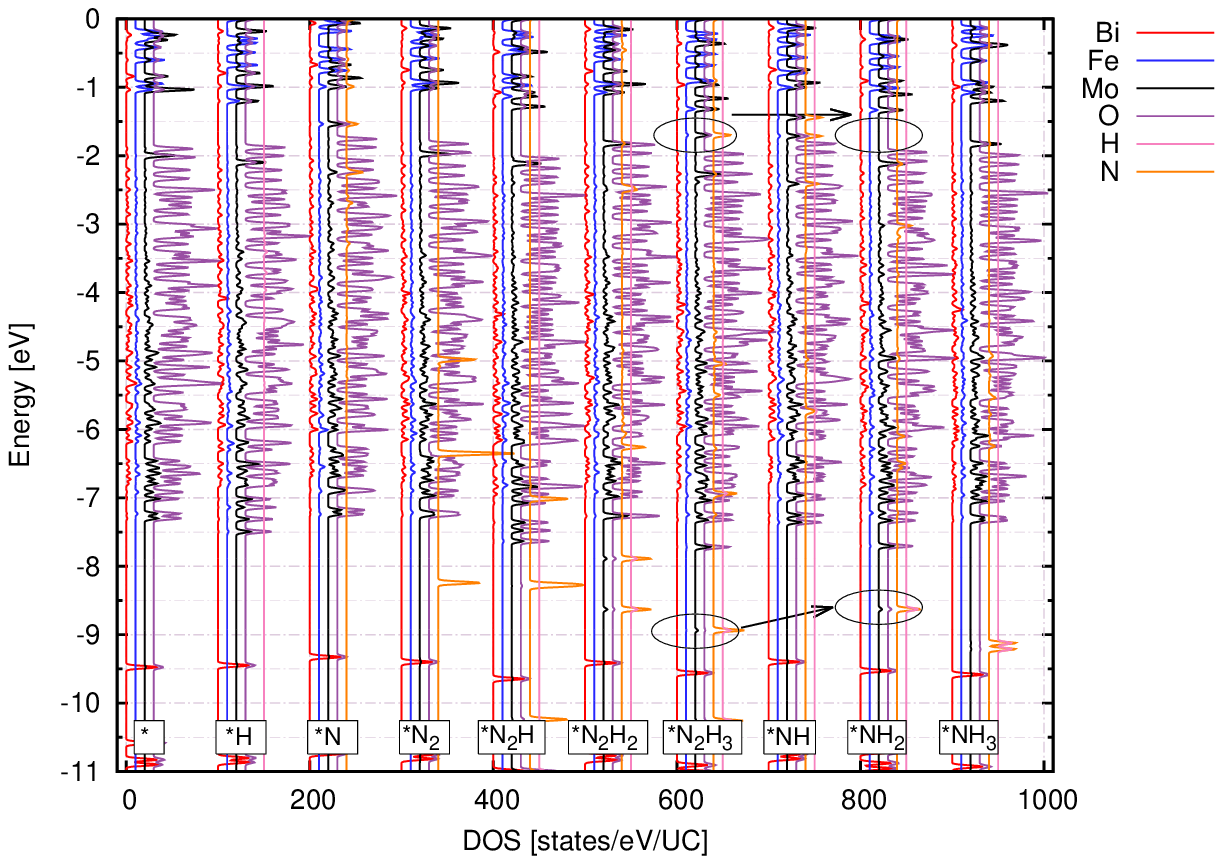}\\
\vspace{-1.8cm}
\begin{flushleft}
\textbf{\begin{Large}B\end{Large}} \\
\end{flushleft}
\vspace{.25cm}
\caption{Plot of the partial density of states (PDOS) for the orthorhombic configuration of (A) Bi$_{2}$MoO$_6$ and (B) (Bi$_{6/8}$Fe$_{2/8}$)$_2$MoO$_6$. The Fermi level has been aligned to 0 eV. Comparison between A and B illustrated the influence that the Fe ion has on the Mo-O bond state, which results in a modulation of the HOMO states. Moreover, the circles in both A and B illustrate the limiting step ($\Delta$G8), where the incorporation of Fe eliminates the covalent bond of NH$_2$ with Mo, reducing the associated reaction barrier. }\hrule
\label{fig:pdos}
\end{figure}

Figure~\ref{fig:pdos} expresses the partial density of states (PDOS) for orthorhombic configuration of A) Bi$_{2}$MoO$_6$ and B) (Bi$_{6/8}$Fe$_{2/8}$)$_2$ MoO$_6$ for all N and H cases implemented into the configuration. The PDOS plot provides information of how the states associated with each atom are interacting with each other. In Figure~\ref{fig:pdos} the Fermi levels have been aligned (set to zero eV) across the plot. All states below the Fermi level are assumed bound. Because the density is determined from the expectation value of the wave function, it is reasonable to assume that atoms with density peaks at similar energy levels have complimentary wave functions. These complimentary wave functions are associated with strong covalent bonds. While the directionality of the covelent bond is desirable, the strength of bond often leads to a thermodynamic limiting step in the case of surface bound states. In addition surface bound states, the PDOS also provide information about the binding within the bulk material. As seen in Figure~\ref{fig:pdos}, there is a modulation of the Mo-O with and without Fe that induces strain. This in turn shifts the energy level and modulates the band gap energy as support by the band gap as a function of composition shown in Figure~\ref{fig:dos}. Furthermore, in Figure~\ref{fig:pdos}, there is a noticeable modification of the Mo and O states as Fe is introduced into the orthorhombic configuration (Figure~\ref{fig:pdos}B) when compared to a pure Bi case (Figure~\ref{fig:pdos}A). This behavior of modifying the Mo-O bond is confirmed for Fe implemented in the structure, as Fe-O will more strongly bond (orbital hybridization) as seen in the PDOS for Fe-O compared to Bi-O. This is seen as the alignment of the Fe states with the O states (strong orbital interaction), note that Fe in Figure~\ref{fig:pdos}B shifts all states up toward the Fermi, thus modifying the composition of the structure with Fe will modify and cause strong covalent bonding. Note that Fe ion causes a considerable modulation of the HOMO Mo and O states, this is apparent in all cases in Figure~\ref{fig:pdos}B. Also, this stronger bond length between Fe-O compared to Yb-O and Bi-O is to be expected as a result of a lower atomic number of Fe. Note that this increased bonding is the cause for the change in the states nearest the HOMO level as seen in Figure~\ref{fig:pdos}B. Also, the contribution of Mo is further validated as being the absorption site for all N and H species, this is seen as alignment of the individual states to Mo states as seen in Figure~\ref{fig:pdos}A,B.

More interesting, Figure~\ref{fig:pdos} provides information about how the surface species bind with the bulk surface. The circles in Figure~\ref{fig:pdos} highlight the transition of the most difficult step, which is associated with $\Delta$G8. Comparing between subfigure A and B where Fe is incorporated there is a clear difference. In the absence of Fe the N-Mo bond shift up as a result of the change in orientation relative to the surface (see Figure~\ref{fig:ads}), which is desirable from an energetic point of view but the associated number of bonds (number of states/energy level) at that energy level increases.  Meaning that NH$_2$ has increased binding resulting in a larger thermodynamic barrier. However, the most intriguing finding comparing between subfigure A and B is that when Fe is incorporated the N-Mo covalent state is not present. And a lower N-Mo state is shifted up in energy. Because the product of the reaction (NH$_2$) is slightly less stable the Gibbs energy ends up being positive, as shown previously. Moreover, as reasoned earlier by measuring the bond lengths, by introducing Fe into the structure we can confirm that from Figure~\ref{fig:pdos}b that there is increased Mo-Fe binding, which modulates the covalent bonds of N-Mo.

The monoclinic configuration was the least desirable of the two configuration due to the limited affinity for nitrogen. This is attributed partially because the Mo-O atoms were not located at the free surface of the slab. It was determined that there were no high index planes which have the Mo on the surface as was the case in the orthorhombic configuration. In turn this did not permit the Mo-O atoms to interact very well with N species. Thus for the monoclinic configurations, all determinant steps of Figure~\ref{fig:max_step}C,D for the associative and dissociative mechanisms have the breaking of N-N as the highest to overcome as seen in Figure~\ref{fig:step}C,D ${\Delta}$G8 and ${\Delta}$G14. For the monoclinic configurations, the two scenarios that assist in explaining what is the better structure for each mechanism are, scenario (i) that for the dissociated mechanism the structure that proves more ideal is the one that has a good balance of N and H species affinity. This is because more energy will be required to break the N-N bond if the structure has a higher affinity for N species than H species. Thus the high combination of Fe, La, Bi, Yb, will tend to have a higher affinity for N species than H species in the monoclinic configuration, and will not prove suitable as seen in Figure~\ref{fig:max_step}C. (Bi$_{7/8}$Yb$_{1/8}$)$_2$MoO$_6$ of the monoclinic dissociated mechanism had the right balance of N and H affinity that results in approximately 7.8 eV as seen for determinant step in Figure~\ref{fig:max_step}C for ${\Delta}$G14 associated with N-N bond breaking. Scenario (ii), for the associative mechanism of monoclinic structure, the configurations that produce better affinity of N species on the surface result in lower determining steps as seen in Figure~\ref{fig:step}D. This is reasoned because continuous protonation steps are required for the associative mechanism before any breaking of the N-N bond. Thus structures in the monoclinic configuration of the associative mechanism that demonstrate better stabilize of N species on the surface to be more effective. This better stability of N species on the surface allows each protonation step to more strongly affect the N-N bond breaking. (Bi$_{6/8}$Fe$_{2/8}$)$_2$MoO$_6$ for the monoclinic configurations as seen in Figure~\ref{fig:step}D to result in a determinant step of approximately 1.7 eV as shown Figure~\ref{fig:max_step}D for N-N bond breaking associated with ${\Delta}$G8.

\section{Volcano Plot}

\begin{figure}[!h]
\includegraphics[width=1\columnwidth]{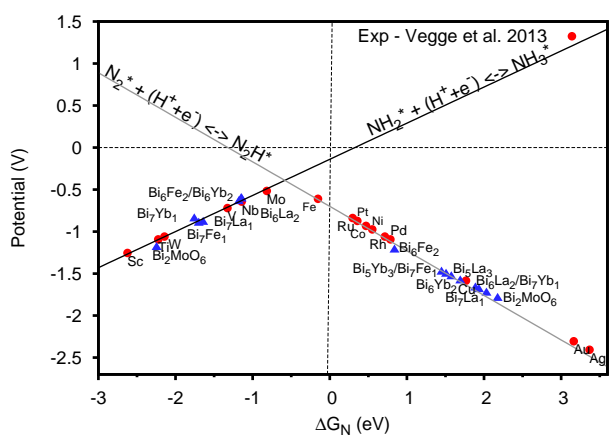} 
\caption{Volcano plot which compares the associative limiting steps for different experimental~\cite{howalt13} (red points) and investigated materials (blue points). Illustrates the volcano plot for the associative mechanism with free energy of ${\Delta}$G$_{N*}$ as the descriptor. The two solid lines indicated in the plot are the limiting step for the associative mechanism, reference~\cite{howalt13} for experimental results.}\hrule
\label{fig:val}
\end{figure}
Figure~\ref{fig:val} illustrates a volcano plot for the associative mechanism. The volcano plot allows for a visual representation of the onset potential required to evolve nitrogen fixation, this method is shown in reference~\cite{howalt13}. In this method the reaction free energy, described by reaction steps corresponding to Equations~\ref{equ:4}-\ref{equ:22}, are used as a simple measure of the electrocatalytic activity. The reaction free energy for each step is expressed as a function of an applied bias. The idea is to find the required bias needed such that each forward reaction has a negative free energy that will make the limiting step zero, which ensures a spontaneous reaction~\cite{howalt13}.

Figure~\ref{fig:val} describes the required bias needed to make each reaction step exothermic. The two solid lines in Figure~\ref{fig:val} shows the limiting steps for the associative electrochemical ammonia synthesis. The red point express material systems experimentally tested in reference~\cite{howalt13} for the two limiting steps and the blue points are the structure configurations theoretically calculated for the orthorhombic bismuth structure system. Where the two solid lines meet is the optimal point where the lowest onset potential for driving associative ammonia synthesis. Note that none of the investigated materials are at the lowest onset potential for driving associative ammonia synthesis electrochemically. However, higher concentration of Fe, La, and Yb seem to perform better at reaching the optimum point. Note that higher concentration will yield stability issues as discussed in earlier section of this chapter, and thus the optimal point may not be achieved with the bismuth based material. Thus finding the optimum material system for the associative mechanism will lie between the Mo and Fe points illustrated in Figure~\ref{fig:val}, which is known to fit quite well with the active site in the nitrogenase enzyme of FeMo cofactor that drives BNF.

\begin{table}
\footnotesize
\begin{center}
\begin{tabular}{ c |c |c }
\hline
 Pathway & Orthorhombic ($\eta$[eV]) & Monoclinic ($\eta$[eV]) \\
 \hline
 Associative & (Bi$_{6/8}$Fe$_{2/8}$)$_2$MoO$_6$ (1.4) & (Bi$_{6/8}$Fe$_{2/8}$)$_2$MoO$_6$ (1.7) \\
 Dissociative & (Bi$_{6/8}$Fe$_{2/8}$)$_2$MoO$_6$ (3.2) & (Bi$_{7/8}$Yb$_{1/8}$)$_2$MoO$_6$ (7.8) \\
 \hline
\end{tabular}
\end{center}
\caption{The best possible compositions for each reaction pathway of the associative and dissociative mechanism of Bismuth bases structures of Orthorhombic and Monoclinic configurations. These correspond to determinant steps attained from Figure~\ref{fig:max_step}. The best structure is that of the orthorhombic Fe containing associative pathway for (Bi$_{6/8}$Fe$_{2/8}$)$_2$MoO$_6$ structure.}
\label{table:best_points}
\end{table}

A summary of the best structure configurations for this study is shown in Table~\ref{table:best_points} with the respected theoretical over-potential calculated for both the associated and dissociated reaction mechanism. It is noted that all expressed over-potential energies are within the band gap energy for the material as expressed in Table~\ref{table:band_gap} except (Bi$_{7/8}$Yb$_{1/8}$)$_2$MoO$_6$ of the monoclinic configuration for the dissociative mechanism, which resulted in an over-potential of approximately 7.8 eV. It should be noted that the closer the over-potential is to the band gap the better the prominence of the overall structure will be, due to more utilization of the electromagnetic spectrum for photocatalysis. Also, the reader should note that the approximated results throughout this study by means of DFT should not be used as exact results, because of the approximations of the DFT method and the expressed approximations should only be used to demonstrate the relevant trends in this study.

\section{Conclusion}
This study implemented DFT and experimental methods to explore design space of (Bi$_{x}$M$_{y}$)$_2$MoO$_6$ where (M = Fe, La, Yb) in both the orthorhombic and monoclinic configurations. The first principle approach allowed understanding the additional physics that would be difficult to determine and express experimentally for this large design space. Investigations are done on the energy gaps and thermodynamic energy barriers for the associative and dissociative reaction pathways that demonstrate strong affinity of hydrogen and nitrogen towards Mo sites. However, all configurations had the same concentration of Mo sites, thus the affinity of N and H species are explored for the orthorhombic and monoclinic substitution of Fe, La, and Yb. Thus this study expresses the balance between the affinity of N and H species for configuration compositions of La, Yb, and Fe that resulted in an orthorhombic structure that fallowed the reaction pathway of the associated mechanism to yield the lowest energy barrier.

Physical insight is gathered through interpretation of bound electronic states at the surface. Compositional phases of higher Fe and Yb concentrations resulted in decreased Mo-O binding and increased affinity between Mo and the N and H species on the surface. The modulation of the Mo-O binding is induced by strain as Yb and Fe are implemented, this, in turn, shifts energy levels and modulates the band gap energy by approximately 0.2 eV.  This modification of Mo-O bond as substitution occurs is a result of the orbital hybridization of M-O  (M = Fe, Yb) that causes a strong orbital interaction that shifts states up toward the Fermi. Thus modifying the composition of the structure will modify and cause strong covalent bonding that causes a considerable modulation of the HOMO Mo and O states. This modification is more apparent for the stronger bond length between Fe-O compared to Yb-O and Bi-O as a result of a lower atomic number of Fe. Note that this increased bonding is the cause of the change in the states nearest the HOMO level. Also, the contribution of Mo is validated as being the absorption site for all N and H species.

Thus the resulting structure was found to be a combination of Bi and Fe in the orthorhombic configuration that tended to have both affinity properties of N and H species in the associative mechanism that is (Bi$_{6/8}$Fe$_{2/8}$)$_2$MoO$_6$ optimal structure. The approximate over-potential was 1.4 eV that is well within the band gap for orthorhombic configuration. This composition demonstrates effective nitrogen and hydrogen affinity that follows the associative or biological nitrogen fixation pathway. Also, the same composition proved ideal for the dissociative pathway that resulted in 3.2 eV for predicted maximum thermodynamic energy barrier.

\section*{Data Availability}
The raw data required to reproduce these findings are available to download by contacting the corresponding author. The processed data required to reproduce these findings are available to download from contacting the corresponding author.

\section*{Acknowledgments}
The authors thank the Super Computing System (Spruce Knob) at WVU, which is funded in part by National Science Foundation EPSCoR Research Infrastructure Improvement Cooperative Agreement \#1003907. A.Y. would like to acknowledge the partial support form DOE ARPA-E  grant number DE-AR0000608.

\footnotesize{
\bibliography{journal} 
\bibliographystyle{rsc} 
}


\end{document}